\documentclass[reprint,aps,prb,english,superscriptaddress]{revtex4-2}

\usepackage{graphicx}
\usepackage{bm,amssymb,amsmath,mathrsfs,amstext,latexsym}

\usepackage[usenames,dvipsnames]{xcolor}
\usepackage[colorlinks=true,citecolor=blue,urlcolor=blue,linkcolor=blue]{hyperref}
\usepackage[normalem]{ulem}
\usepackage{booktabs}
\newcommand{\der}[2]{\frac{d #1}{d #2}}

\let\tilde\relax
\newcommand{\tilde}[1]{\widetilde{#1}}

\begin{document}

\title{Renormalization-Group Analysis of the Many-Body Localization Transition in the Random-Field XXZ Chain}

\author{Jacopo Niedda}
\email{jniedda@ictp.it}
\affiliation{The Abdus Salam ICTP, Strada Costiera 11, 34151 Trieste, Italy} 

\author{Giacomo Bracci Testasecca}
\email{gbraccit@sissa.it}
\affiliation{SISSA, via Bonomea 265, 34136, Trieste, Italy}
\affiliation{INFN, Sezione di Trieste, Via Valerio 2, 34127 Trieste, Italy}

\author{\\Giuseppe Magnifico}
\affiliation{Dipartimento di Fisica, Università di Bari, I-70126 Bari, Italy}
\affiliation{Istituto Nazionale di Fisica Nucleare (INFN), Sezione di Bari, I-70125 Bari, Italy}

\author{Federico Balducci}
\affiliation{Max Planck Institute for the Physics of Complex Systems, N\"othnitzer Str. 38, 01187 Dresden, Germany}

\author{Carlo Vanoni}
\affiliation{SISSA, via Bonomea 265, 34136, Trieste, Italy}
\affiliation{Department of Physics, Princeton University, Princeton, New Jersey, 08544, USA}

\author{Antonello Scardicchio}
\affiliation{The Abdus Salam ICTP, Strada Costiera 11, 34151 Trieste, Italy}
\affiliation{INFN, Sezione di Trieste, Via Valerio 2, 34127 Trieste, Italy}

\date{\today}


\begin{abstract}
    The spectral properties of the Heisenberg spin-1/2 chain with random fields are analyzed in light of recent works on the renormalization-group flow of the Anderson model in infinite dimension. We reconstruct the beta function of the order parameter from the numerical data, and observe that it may not admit a one-parameter scaling form and a simple Wilson-Fisher fixed point. Rather, it appears to be more compatible with a two-parameter, Berezinskii–Kosterlitz-Thouless-like flow with a line of fixed points (the many-body localized phase) terminating at the localization transition critical point. We argue that this renormalization group framework provides a more coherent and intuitive explanation of numerical data, up to the system sizes available with the present technology.
\end{abstract}

\maketitle


\section{Introduction}

Dynamical properties of disordered systems have been at the center of an intense activity of research since the 1950s. Surprising behavior and new physics has been found in the dynamics of spin glasses~\cite{edwards1975theory,binder1986spin,mezard1987spin}, and of quantum particles in a random potential, where the most counter-intuitive effect is the complete localization of such particles at large disorder~\cite{Anderson1958absence,evers2008anderson}.

Despite the enormous amount of progress on the single-particle Anderson localization, including experiments with cold atoms in optical lattices \cite{roati2008anderson,billy2008direct}, the problem of whether localization persists in an ensemble of {\it weakly interacting} particles has traditionally been a source of controversy. In the early 1980s, perturbative calculations pointed to the resilience of the localized phase towards the introduction of interactions~\cite{fleishman1980interactions}, and a series of papers bootstrapping perturbative calculations by means of renormalization group (RG) ideas appeared~\cite{finkelstein_83,castellani_1984,castellani_1984_2}. 
The difficulty of the problem appeared soon, since no controlled calculation or numerical experiment was considered resolutive of the situation. The question of whether an arbitrarily small interaction destroys localization or not has remained open, and for some, it still is \cite{sierant24MBLreview}. 

In recent years, the topic has seen a resurgence of interest mainly due to the appearance of the works by Gornyi, Mirlin, and Polyakov~\cite{Gornyi2005Interacting} and by Basko, Aleiner, and Altshuler (henceforth BAA)~\cite{Basko06}; the latter addressed the problem through the calculation of quasiparticle lifetime to all orders in perturbation theory. 
To tame such a formidable task, BAA necessarily resorted to some approximation, which allowed them to map the problem to a sort of Anderson model on a tree with correlated energies, in the spirit of Ref.~\cite{altshuler1997quasiparticle}. However, they also incorporated the spatial dimension $d$, with the connectivity representing the number of \emph{physical channels} open for transmission at a given energy. The result of the calculation is a critical value for the interaction strength, below which the localized quasiparticle lifetime is typically infinite. This phase is called many-body localized (MBL).

Some time after the BAA paper, people started investigating the problem using exact diagonalization of small systems~\cite{oganesyan2007localization,Znidaric2008Many,Pal10,Bardarson2012Unbounded,Deluca13,Luitz15,Pietracaprina2018Shift,Sierant2020Polynomially,Colbois2024prl,colbois2024b} and the concept of local integrals of motion (LIOMs) arose~\cite{serbyn2013local,Huse2014Phenomenology}. The existence of LIOMs in the MBL phase was proved, within the same approximations used in BAA, in Ros {\it et al}~\cite{ros2015integrals}, but it is still debated whether those approximations are crucial for the result or can be justified. The question, therefore, is whether the neglected diagrams in BAA, which deform the tree-like geometry into the real Fock-space geometry underlying the perturbation theory, would change the renormalization group equations to the point of eliminating the transition altogether. That this is not the case in one dimension, and under sufficiently weak interactions, is the claim of Refs.~\cite{Imbrie2016Many,Imbrie2016Diagonalization} (see also the review~\cite{Imbrie17}). Those claims, however, whose reliance on some additional assumptions on the statistical behavior of energy levels has been contested, have been recently revisited and found to be sound~\cite{deroeck2024absence}.

In this work, we take a different approach to analyzing the numerical data obtained by exact diagonalization of small systems, using the numerical data and the theory of finite-size scaling to infer the RG flow of the MBL transition in the light of the recent works \cite{altshuler2024renormalization,vanoni2023renormalization}. Within the accuracy of our analysis, the numerical reconstruction of the model phase diagram appears consistent with the picture already discussed for the Anderson model in infinite dimension \cite{vanoni2023renormalization}: the ergodic fixed point has the nature of a random matrix theory (RMT) fixed point~\cite{kutlin2024investigating}, with volumic scaling (i.e.\ the natural scaling variable is the Hilbert space volume and not the physical volume of the system), while the critical point seems to lie at the end of a line of fixed points representing the localized phase.

The question is then reduced to understanding the flow near the line of fixed points, which is qualitatively reminiscent of a Berezinskii–Kosterlitz-Thouless (BKT) RG flow (the appearance of BKT physics is a recurring theme in MBL \cite{Goremykina2019Analytically,Dumitrescu2019Kosterlitz,Morningstar2019Renormalization,Morningstar2020Many}). We show that the flow can have two possible topologies. In one, like in the BKT analysis of the XY model or the Anderson model in infinite dimensions, the line of fixed points is attractive and corresponds to the MBL phase. Alternatively, the line can be repulsive, and the MBL phase does not exist for any value, no matter how large, of disorder. To discriminate between the two scenarios, we analyze the numerical data at moderately high disorder in terms of the RG trajectories. We further realize that the RG flow can be mapped onto the phase space orbit of a classical, Newtonian particle. We try to infer the potential in which the particle moves, and from this, the fate of the MBL phase. \medskip

The paper is structured as follows. In Sec.~\ref{sec:model} we introduce the model studied and the observable used in order to probe the presence of a disorder induced localization transition. In Sec.~\ref{sec:scaling_theory} we review the renormalization group approach based on the numerical study of the scaling theory of localization, highlighting the difference between the standard scenario of one-parameter scaling hypothesis and the two-parameter scaling hypothesis. Furthermore, we discuss a family of dynamical systems that describe the RG flow, and we show how this framework allows for a better interpretation of numerical data. In Sec.~\ref{sec:numerical_results} we discuss the numerical data and the problem of the significance of the statistical ensemble used; finally, we draw our conclusions and discuss future perspectives in Sec.~\ref{sec:conclusions}.

\section{Model}
\label{sec:model}

The random-field XXZ spin chain is defined by the Hamiltonian
\begin{equation}
    \label{eq:H_XXZ}
    \hat{H} = J \sum_{i=1}^L   \hat{\mathbf{S}}_i \cdot\hat{\mathbf{S}}_{i+1} + \sum_{i=1}^L  h_i \hat{S}_i^z  \; ,
\end{equation}
where $\hat{\mathbf{S}}_i = (\hat{S}_i^x, \hat{S}_i^y, \hat{S}_i^z) $ are spin-$1/2$ operators and $h_i$ are i.i.d.\ random variables distributed uniformly over the interval $[-W,W]$. Since the relevant parameter is the ratio between the disorder strength $W$ and the hopping parameter $J$, in the following we will fix $J=1$ and consider $W$ as the only parameter. The model has a global symmetry given by the conservation of the total spin $\hat{S}^z=\sum_i \hat{S}^z_i $. For numerical efficiency, in this work we will always restrict to the $\hat{S}^z = 0$ sector, whose dimension is given by 
$ {\rm dim (\mathcal{H})} = \binom{L}{L/2} \lesssim  2^L.$ 

It has been conjectured that the Hamiltonian~\eqref{eq:H_XXZ} undergoes a many-body localization phase transition at a critical disorder strength $W_c$~\cite{Znidaric2008Many,Pal10}: while a small disorder $W$ breaks the Bethe-ansatz integrability of the clean model and makes it ergodic, for $W>W_c$ the system becomes a non-ergodic insulator, failing to thermalize.
Moreover, studies of transport properties in the disordered Heisenberg chain showed the presence of a transition from diffusive to subdiffusive transport within the ergodic phase~\cite{vznidarivc2016diffusive,Diggen2018Manybody}.
It would be impossible to summarize in few lines the huge body of work that have investigated the features of this intriguing non-thermal phase of matter, and of the phase transition leading to it: we refer the reader to the many reviews~\cite{Altman2015Universal,Nandkishore2015Many,Alet2018Many,Abanin2019colloquium,Gopalakrishnan2020Dynamics,sierant24MBLreview}. However, here we stress the main points, that lead to the still open questions in the field. 

On one hand, the perturbative computation leading to MBL~\cite{Basko06}, extended in Refs.~\cite{ros2015integrals,Imbrie2016Diagonalization,*Imbrie2016Many,Imbrie2016Local} to spin chains, predicts that the eigenstates at strong enough disorder assume the form of exponentially-localized integrals of motion (LIOMs)~\cite{serbyn2013local,Huse2014Phenomenology}. Correspondingly, the spectral statistics is Poissonian. On the other hand, it has been argued that non-perturbative effects such as avalanches~\cite{DeRoeck2017Stability,Thiery2018Many} and many-body resonances~\cite{Ha2023Manybody,Morningstar2022Avalanches}, can lead to delocalization. The conflict between these opposite views seems not to be solvable by numerical methods, as exact diagonalization is confined to too small systems~\cite{Chandran2015Finite,panda2020can,abanin2021distinguishing}, and tensor network methods to too short times~\cite{sierant2022challenges}. The absence of clear-cut numerical results has led different authors to claim that MBL does not exist at all~\cite{Suntajs2021Spectral,Sels2021Dynamical} or that a wide, pre-thermal regime hinders the numerical observation of the actual MBL transition~\cite{Morningstar2022Avalanches,Crowley2022Constructive}.


In modern studies of disorder-induced localization in quantum systems, it is customary to analyze the spectral properties to probe the ergodic or localized nature of the model. A quantity typically considered for discriminating between chaotic or localized spectra is the spectral gap ratio, or $r$-parameter~\cite{oganesyan2007localization}:
\begin{equation}
    \label{eq:r_L_def}
   r=\frac{1}{N_e-2}\sum_{n=1}^{N_e-2}\frac{\min(\Delta E_n,\Delta E_{n+1})}{\max(\Delta E_n,\Delta E_{n+1})},
\end{equation}
where $\Delta E_n = E_{n+1}-E_{n}$ is the $n$-the energy gap in a narrow energy window containing $N_e$ energy levels. Notice that the use of such a window in the ergodic case would define the microcanonical ensemble. The value $r_{\mathrm{P}} = 2 \ln 2 - 1 \simeq 0.386$ is the Poisson value for integrable systems, while $r_{\mathrm{WD}} \simeq 0.5307$ is the Wigner-Dyson value for random matrices of the Gaussian Orthogonal Ensemble (GOE), corresponding to chaotic systems with time-reversal symmetry. For convenience, we rescale the average gap ratio, defining 
\begin{equation}
    \label{eq:def_phi}
    \phi = \frac{r - r_{\mathrm{P}}}{r_{\mathrm{WD}}- r_{\mathrm{P}}}.
\end{equation}
It follows that $\phi = 0$ corresponds to a non-ergodic behaviour, while $\phi=1$ to ergodicity. 

\begin{figure}
    \centering
    \includegraphics[width=\columnwidth]{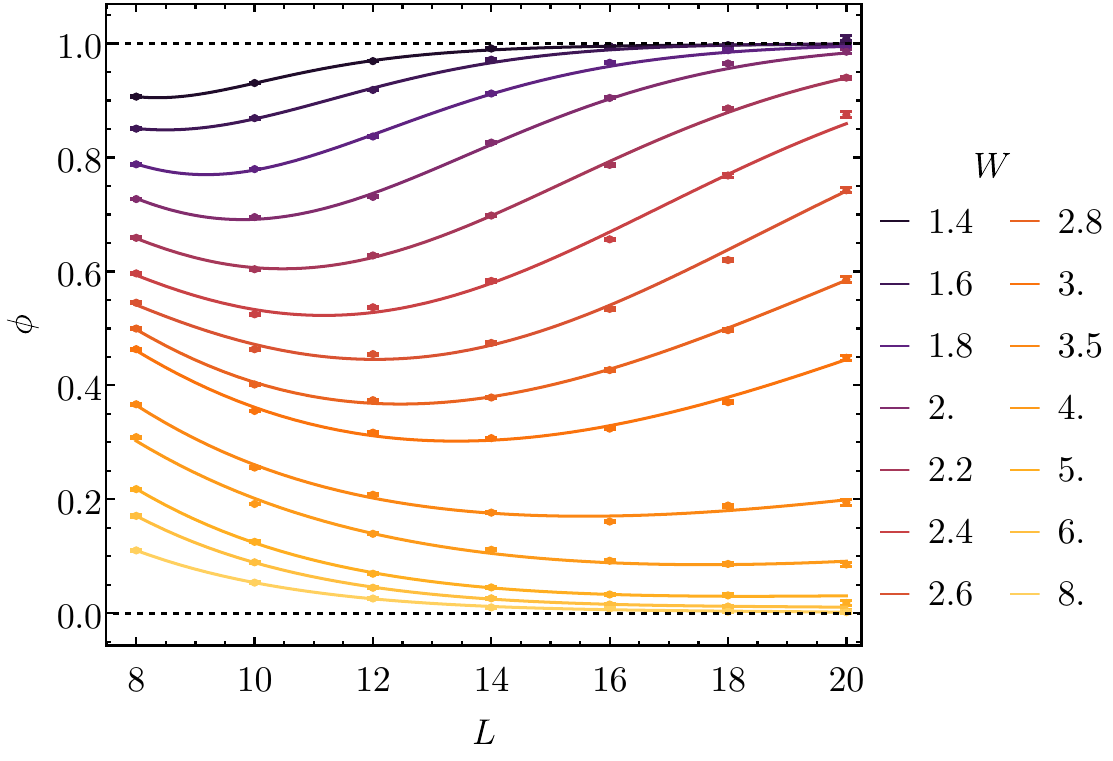}
    \caption{System-size dependence of the rescaled $r$-parameter $\phi$, Eq.~\eqref{eq:def_phi}, in the XXZ spin chain. Different colors correspond to different values of the disorder strength $W$. Dots represent numerical data, while solid lines are polynomial interpolating curves, later used for the $\beta$-function computations. Dashed lines represent the asymptotic values: $\phi=0$ for the localized and $\phi=1$ for the ergodic phase. }
    \label{fig:XXZ-rpar}
\end{figure}

The rescaled $r$-parameter of the random-field XXZ spin chain averaged over many disordered samples is shown as a function of the system size in Fig.~\ref{fig:XXZ-rpar}. The data is obtained via exact diagonalization, see App.~\ref{app:sec:EDspec} for a short summary of the numerical methods (see also Ref.~\cite{Pietracaprina2018Shift} for an extended discussion). The solid lines are polynomial interpolations of the data; we comment extensively on the numerical analysis in Sec.~\ref{sec:numerical_results}. First, however, we introduce in the next section the framework of the beta function, necessary to interpret coherently the numerical data.

\section{Scaling theory of localization}
\label{sec:scaling_theory}

In this section, we briefly review the renormalization group (RG) approach, based on the construction of the beta function from numerical data, for the scaling theory of localization. 

The use of RG to study localization transitions goes back to the seminal work of Abrahams, Anderson, Licciardello, and Ramakrishnan~\cite{abrahams1979scaling}. The great achievement of this ``gang of four'' was to understand that a one-parameter scaling hypothesis explains the localization of non-interacting particles in $d \leq 2$ for any disorder strength, and the presence of a finite-disorder localization transition in $d\geq 3$.

Extending the RG analysis to interacting models, however, has proven to be difficult. Some authors took inspiration from the Ma-Dasgupta-Hu-Fisher strong-disorder RG~\cite{Ma1979Random,Dasgupta1980Low,Fisher1992Random,*Fisher1994Random,*Fisher1995Critical}, and applied a similar procedure to models hosting MBL~\cite{Vosk2013Many,Pekker2014Hilbert}. However, more promising results were found upon applying the same real-space RG rules to phenomenological models of conducting/insulating regions~\cite{Vosk2013Many,Potter2015Universal,Zhang2016Many,Thiery2017Microscopically}. In this framework, some evidence for a BKT-like phase transition was found~\cite{Goremykina2019Analytically,Dumitrescu2019Kosterlitz,Morningstar2019Renormalization,Morningstar2020Many}. Here, we set the basis for a controlled RG analysis of a \emph{microscopic} model of MBL.

\subsection{One parameter RG flow}
\label{sec:1PS}

Consider a generic observable $A$---that later will be specified to the rescaled average gap ratio $\phi$---and define its beta function as
\begin{equation}
    \label{eq:1p_scaling}
    \beta \equiv \der{\ln A}{\ln N},
\end{equation}
where $N$ is the volume of the system. One can identify the volume either with the Hilbert space dimension $N \sim 2^L$, or with the spatial volume $N=L^d$. In the definition of the beta function, we have traded the RG parameter for the system size, which in the context of finite-size scaling can be thought of as a relevant operator of the renormalization group~\cite{Cardy1988,PELISSETTO_RG}. We will use the convention $A \in [0,1]$, with asymptotic values $A_\mathrm{loc}=0$ for the localized fixed point and $A_\mathrm{erg}=1$ for the ergodic one, which indeed agrees with the rescaled $r$-parameter $\phi$.

If the one-parameter scaling hypothesis~\cite{abrahams1979scaling} holds, the beta function is a function of $A$ alone, i.e.
\begin{equation}
    \der{\ln A }{\ln N}=\beta(A).
\end{equation}
Therefore, the dependence of $A$ on the system size $\ln N$ is obtained by solving the differential equation above by separation of variables:
\begin{equation}
\label{eq:scaling_sep_var}
    \int_{A_0}^A\frac{dA'}{A'\beta(A')}=\ln \frac{N}{N_0},
\end{equation}
where $A_0$ is the value of $A$ at $N=N_0$. If one is far from a zero of the beta function, one can integrate formally and find 
\begin{equation}
    G(A)-G(A_0)=\ln\frac{N}{N_0},
\end{equation}
where $G$ is a primitive of the integral in Eq.~\eqref{eq:scaling_sep_var}. Inverting this equation for $A$, one finds
\begin{equation}
    A=f(N/\tilde N_0),
\end{equation}
where $f(x) = G^{-1}(\ln x)$ turns out to be a universal scaling function, in one-to-one correspondence with the function $\beta_0$, and $\tilde N_0 \equiv N_0 e^{-G(A_0)}$. Therefore, in a regular section of the flow and far from a critical point, the observables are functions of the volume $N$, whether this volume is the spatial or the Hilbert space volume.

\begin{figure}
    \centering
    \includegraphics[width=\linewidth]{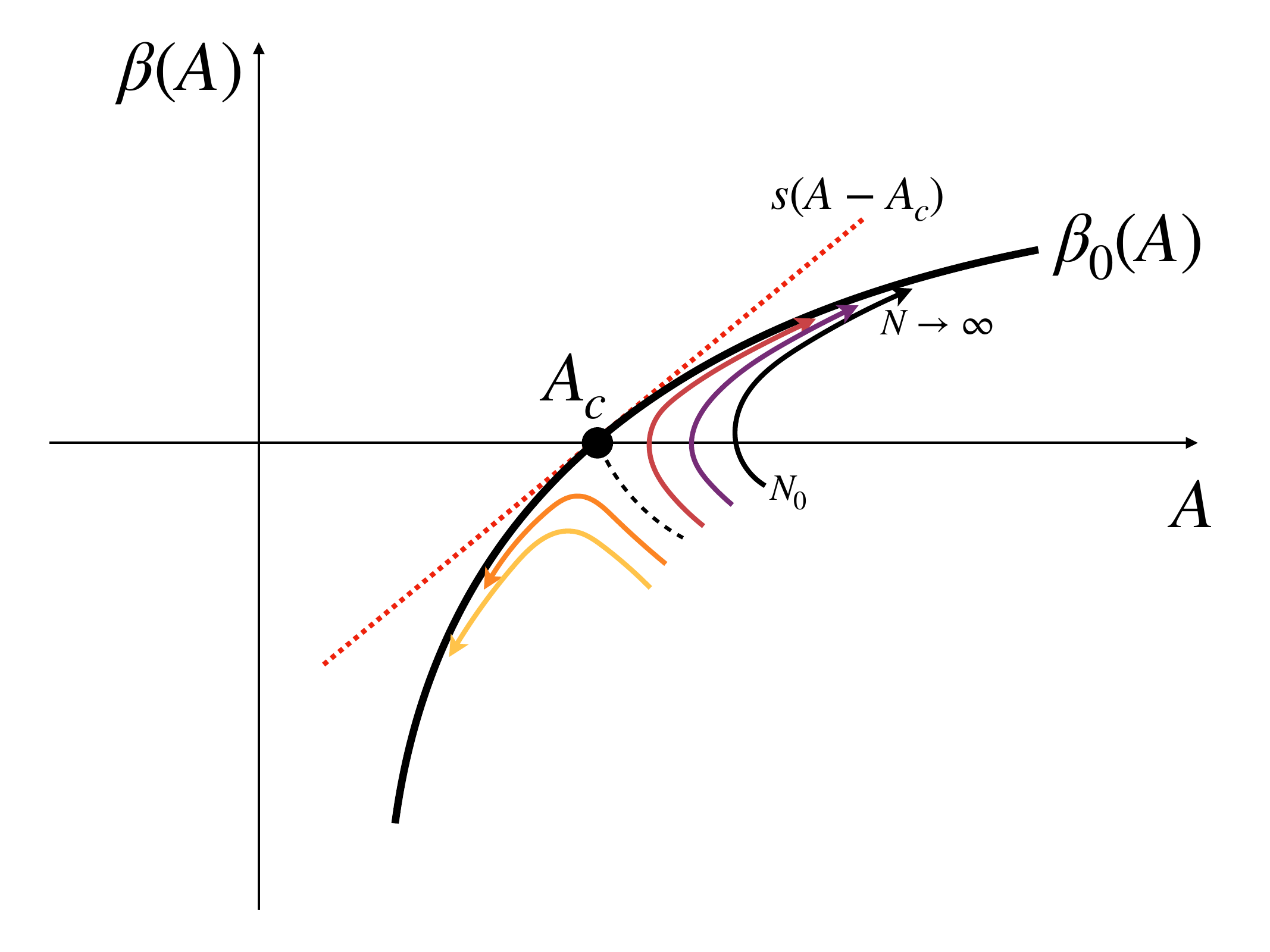}
    \caption{Sketch of a one-parameter-scaling RG flow with irrelevant corrections, similar to what can be observed in the Anderson model in $d$-dimensional space~\cite{altshuler2024renormalization}. The solid black curve represents the one-parameter-scaling beta function $\beta_0(A)$, which vanishes at the critical point $A_c$. The colored solid lines represent the beta function for different choices of the initial parameters, as described by Eq.~\eqref{eq:1PS_irrelevant_corrections}. The starting point of these lines occurs at $N=N_0$ and $N$ grows in the direction indicated by the arrows. Around the critical point $A_c$, the beta function can be linearized as described by Eq.~\eqref{eq:beta_crit_1PS}.}
    \label{fig:sketch_1PS}
\end{figure}

If, instead, the RG trajectories---obtained varying the values of the initial parameters (e.g.\ disorder strength)---do not always lie on a single curve, one needs to take into account corrections to the one-parameter scaling solution. For a fixed value of the initial parameters, let us assume that the curve starts close to the one-parameter curve $\beta_0(A)$
\begin{equation}
    \der{\ln A}{\ln N} \bigg|_{N=N_0}=\beta_0(A)+c \qquad \text{with }c\ll \beta_0,
\end{equation}
and quickly converges to it; see Fig.~\ref{fig:sketch_1PS} for a sketch. One can then write a phenomenological set of equations, projecting the RG flow in the plane spanned by the relevant and the least irrelevant operator
\begin{align}
    \label{eq:2PS-eqs1}
    \der{\ln A}{\ln N} &= \beta, \\
    \label{eq:2PS-eqs2}
    \der{\beta}{\ln N} &=-\omega\big(\beta-\beta_0(A)\big). 
\end{align}
In the limit $\omega\to\infty$, one recovers the one-parameter scaling equation $d\ln A / d \ln N=\beta_0(A)$. Taking $\omega$ large but finite, and assuming to start sufficiently close to the one-parameter scaling curve, such that $\beta_0(A)$ can be considered a constant for a first part of the RG flow, one can find the solution
\begin{equation}
    \der{\ln A}{\ln N}=\beta_0(A)+c\, e^{-\omega\ln (N/N_0)}.
\end{equation}
Upon integration, in the approximation $c\ll 1$, one has
\begin{equation}
    \label{eq:1PS_irrelevant_corrections}
    A(N,A_0)=f(N/\tilde{N}_0)+ f_1(N/\tilde{N}_0) N^{-\omega},
\end{equation}
where $\omega$ corresponds to the dimension of the smallest irrelevant operator. Only rarely $\omega$ is so large that it can be ignored altogether. In particular, for the Anderson model in the limit of $d\to\infty$ --- which corresponds to the model defined on the random regular graph (RRG) --- it has been argued that actually $\omega\to 0$~\cite{altshuler2024renormalization,vanoni2023renormalization}. 
Let us mention that the limit $d\to \infty$ has to be taken with caution and cannot be obtained by simply setting $\omega = 0$ in Eq.~\eqref{eq:1PS_irrelevant_corrections}. We comment more on this in Sec.~\ref{sec:crit_point}.

\subsection{Critical point as a simple zero of $\beta$}
\label{sec:crit_point}

Close to a critical point, if the beta function has a simple zero $\beta(A_c)=0$ for some value $A_c \neq 0$ of the observable, then the dependence of $A$ on $N$ acquires the form of a scaling law. Let us write the Taylor expansion of $\beta(A)$ close to the fixed point as
\begin{equation}
\label{eq:beta_crit_1PS}
    \beta(A)=s(A-A_c)+\sum_{n\geq 2}\frac{c_n}{n!}(A-A_c)^n.
\end{equation}
Going back to Eq.~\eqref{eq:scaling_sep_var}, it holds
\begin{align}
    \int_{A_0}^A\frac{dA'}{A'\beta(A')}
    &=\int_{A_0}^A dA'\frac{1}{A_c s(A'-A_c)} \nonumber \\
    &\hspace{0.5cm} + \int_{A_0}^A dA' \frac{A_c s(A'-A_c)- A'\beta(A')}{A_c s(A'-A_c)A'\beta(A')}\nonumber\\
    &=\frac{1}{A_c s}\ln \left| \frac{A-A_c}{A_0-A_c} \right|+\int_{A_0}^A dA' g(A'-A_c).
\end{align}
The function
\begin{equation}
    g(x)=\frac{A_c s x-(x+A_c)\beta(A_c+x)}{A_c s x(x+A_c)\beta(A_c+x)}
\end{equation}
is regular close to $x=0$:
\begin{equation}
    \label{eq:g_expansion}
    g(x)= -\frac{2s + c_2 A_c}{2A_c^2 s^2} + O(x) \equiv -a + O(x),
\end{equation}
where again $c_n=d^n\beta/dx^n|_{x\to 0}$.

Putting everything together and defining the primitive $\int dx\, g(x)=G(x) = -ax + O(x^2)$ and the critical exponent $\nu=1/ A_c s$, one has
\begin{equation}
    \ln\frac{N}{N_0} = \nu\ln \left|\frac{A-A_c}{A_0-A_c}\right|+G(A-A_c) -G(A_0-A_c),
\end{equation}
which can be inverted first as
\begin{equation}
    |A-A_c|e^{G(A-A_c)/\nu}=\left(\frac{N}{N_0\xi(A_0)}\right)^{1/\nu},
\end{equation}
with $\xi(A_0)=|A_0-A_c|^{-\nu} e^{-G(A_0-A_c)}\simeq |A_0-A_c|^{-\nu}$ being a critical length. Consequently, by calling $f$ the inverse of the function $x \mapsto xe^{G(x)/\nu}$, one gets
\begin{equation}
    \label{eq:1PS-FSS}
    A(N,A_0) = A_c + f \bigg(\bigg(\frac{N}{N_0\xi(A_0)}\bigg)^{1/\nu} \bigg).
\end{equation}
Notice that for small $x$ it holds $xe^{G(x)/\nu}\simeq xe^{- \frac{a}{\nu} x}\simeq x-\frac{a}{\nu}~x^2+O(x^3)$, therefore the function $f(x)\simeq x$ in the same region. 

Under the hypothesis that, at $\ln N_0=O(1)$, $A_0$ is a smooth function of the microscopic parameters of the model, it is possible to collapse the data with a universal scaling function $f$, in one-to-one correspondence with the function $\beta$. In the context of localization transitions, the Anderson model in finite dimensions $d>2$ exhibits a phase transition which is characterized precisely by this kind of scaling theory. Including also the irrelevant corrections discussed in Sec.~\ref{sec:1PS}, one finds
\begin{equation}
    \label{eq:critical_irrelevant_corrections}
    A(N,A_0) = A_c + f(x)  + N^{-\omega } f_1(x),
\end{equation}
where $\omega$ is the scaling dimension of the least irrelevant operator~\footnote{It is possible to work not strictly in the limit $\omega\to\infty$, and analyze the flow close to the critical point to extract the correct relevant and irrelevant anomalous dimensions to $O(1/\omega)$.} and $x \equiv (N/N_0\xi(A_0))^{1/\nu}$. Notice that Eq.~\eqref{eq:critical_irrelevant_corrections} has a notation similar to Eq.~\eqref{eq:1PS_irrelevant_corrections}, but the scaling functions involved are different, as an explicit linearization was performed around the critical point. 

As mentioned previously, in the limit $d \to \infty$ of the Anderson model (corresponding to the RRG), the irrelevant exponent $\omega$ becomes marginal, but this does not correspond to simply setting $\omega = 0$ in Eq.~\eqref{eq:critical_irrelevant_corrections}. In fact, Eq.~\eqref{eq:critical_irrelevant_corrections} has been derived for a critical point $A_c$ isolated from the other “Gaussian" fixed points, while in the RRG $A_c = A_\mathrm{loc} = 0$. One can show that, when these two limits are taken consistently, one obtains logarithmic corrections to Eq.~\eqref{eq:critical_irrelevant_corrections}, indicating the occurrence of a marginal direction~\cite{vanoni2023renormalization}.

Let us see what happens if we collapse our numerical data by working in the framework of the one-parameter scaling. As discussed above, the subleading corrections to Eq.~\eqref{eq:1PS-FSS} are taken into account as in Eq.~\eqref{eq:critical_irrelevant_corrections}. 
Near the critical point, the critical length $\xi(A_0)$ diverges, and therefore $x \ll 1$. In this limit, one can expand $f_1(x) = c + c_1 x + O(x^2)$, and we will consider the non-zero term $f_1(x) = c$, which is a customary choice in the literature (see, {\it e.g.}, Ref.~\cite{sierant2023universality}).
In order to collapse the rescaled $r$-parameter $\phi$, we use the relation
\begin{equation}
    \label{1PS-FSS-phi-corrections}
    \phi(L,W) = \phi_c + f\left(L^{1/\nu} \frac{W-W_c}{W_c} \right) + c L^{-\omega},
\end{equation}
where five constants---$\phi_c$, $W_c$, $\nu$, $\omega$ and $c$---and a function $f$ are to be determined. In the argument of the scaling function, we have replaced $\xi \sim (W-W_c)^{-\nu}$ according to the usual scaling law, and we have chosen to use the chain length $L$ instead of the Hilbert space dimension $N$, since it turned out to provide a better collapse of the data.

\begin{figure}
    \centering
    \includegraphics[width=\columnwidth]{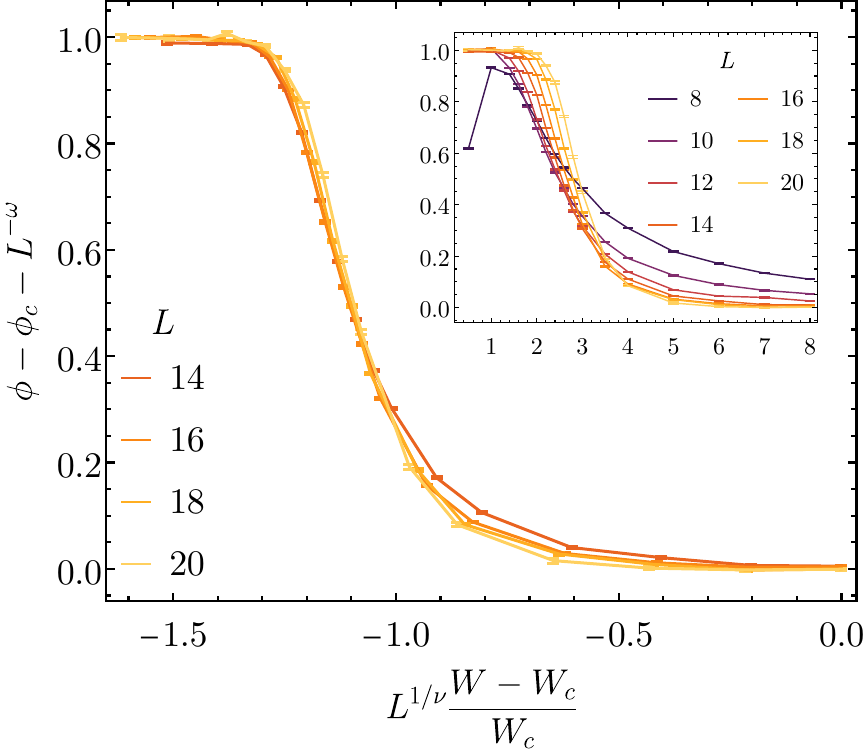}
    \caption{Naive finite-size scaling for the average rescaled gap ratio $\phi$. In the main plot, the collapse is shown for the larger sizes, $L>12$, while the inset contains data before the collapse down to size $L=8$. The collapse is obtained by fixing $\phi_c=0$, and it yields a critical disorder $W_c \simeq 8.0$ and exponents $\nu \simeq 5.5$, $\omega \simeq 2.0$. We will present a different explanation for the critical exponent $\omega=2$ in the next section. Notice also the similarity with the data for the Anderson model on the RRG in \cite{sierant23a}. For a collapse in which we allow $\phi_c$ to vary, we refer to App.~\ref{app:sec:datacollapse}.}
    \label{fig:scalingXXZ}
\end{figure}    

In Fig.~\ref{fig:scalingXXZ} a fairly good data collapse is obtained for the rescaled $r$-parameter by fixing $\phi_c=0$, which is what one expects observing the data in Fig.~\ref{fig:XXZ-rpar}. The collapse yields a critical disorder $W_c \simeq 8.0$ and exponents $\nu \simeq 5.5$ and $\omega \simeq 2.0$. The collapse, however, cannot be taken too seriously. First, such a large $\nu$ exponent seems to signal an exponential scaling with system size near the critical point, rather than a power-law. Second, and more importantly, the function $f$ in Eq.~\eqref{1PS-FSS-phi-corrections} appears not to be linear near $x=0$ in Fig.~\ref{fig:scalingXXZ}, i.e.~$df/dx=0$ at the critical point, and this is a clear contradiction with the previous discussion. 

As a workaround (see App.~\ref{app:sec:datacollapse} for details), one could leave $\phi_c$ free and obtain a fairly good collapse by fixing the critical value of the disorder to $W_c \simeq 3.8$, which can be linearly extrapolated from the finite-size zeros of the beta function (see e.g.\ the points $\phi_A$ in Fig.~\ref{fig:phiA_vs_W} up to $W=3$), and which agrees with the literature~\cite{Luitz15}. However, this collapse yields a critical value $\phi_c \simeq 0.13$, which is larger than the upper bound one finds from the same numerical data: for $W=4$ one already sees a turning point at $\phi_A = 0.085(4)$. Moreover, the best fit critical exponents are $\nu \simeq 1$ and $\omega \simeq 2$, with $\nu$ violating the Harris bound $\nu > 2/d$~\cite{Harris1974Effect,Chandran2015Finite}, where the spatial dimension is $d=1$ in the present case.

\subsection{From one to two parameters}
\label{sec:1to2PS}

\begin{figure*}
    \centering
    \includegraphics[width=0.45\textwidth]{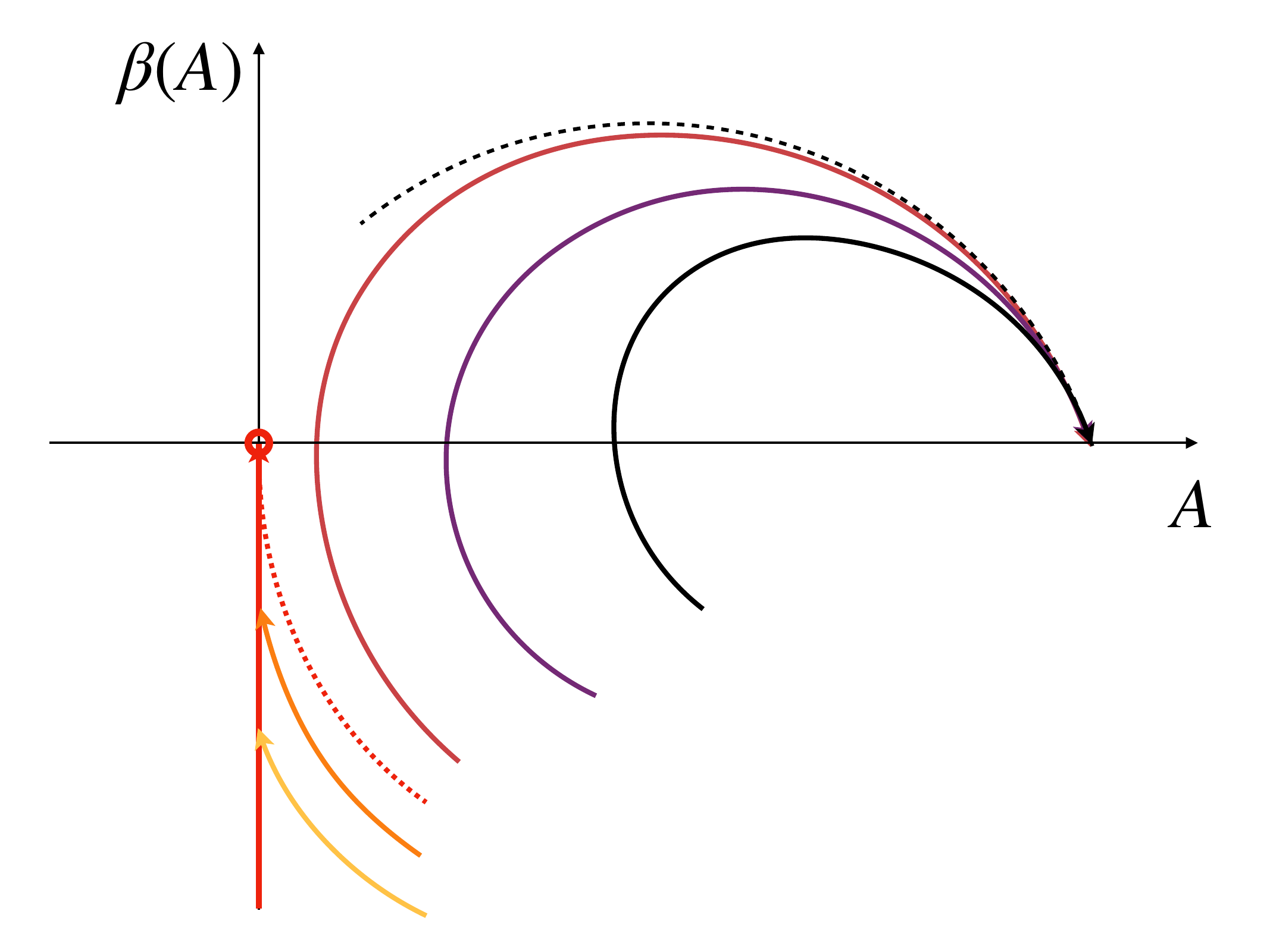}
    \includegraphics[width=0.45\textwidth]{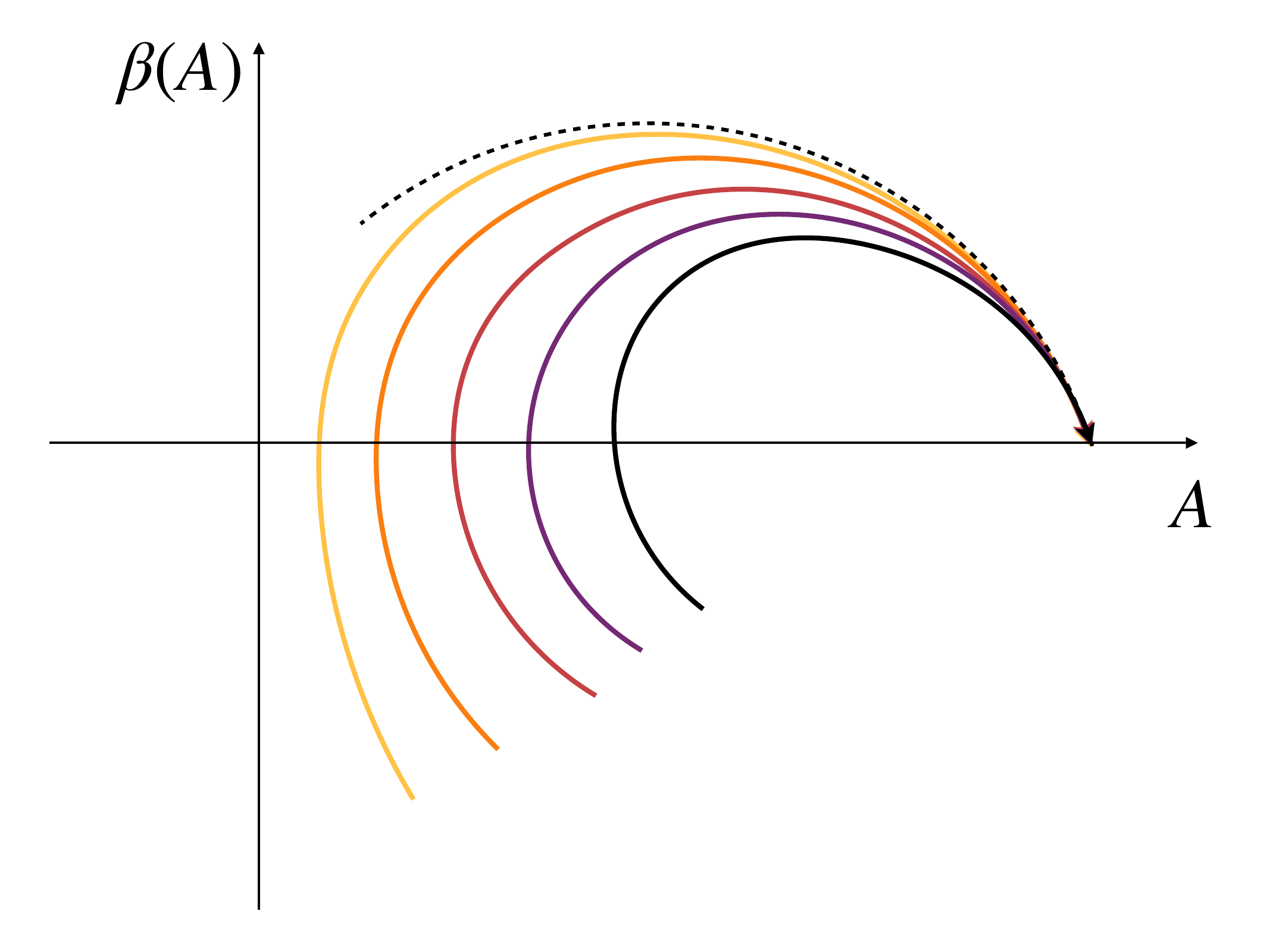}
    \caption{\emph{(Left)} Beta function for a two-parameter scaling transition scenario. In the localized phase, the RG flow reaches $A=0$ at a finite and negative $\beta = -1/\xi_\mathrm{loc}$, and its value depends on $W$, thus generating a line of fixed points (solid red). In the ergodic phase, instead, the curves flow to $A=1$ as prescribed by random matrix theory. The RG flow around the ergodic fixed point is described by a one-parameter scaling, meaning that all the RG curves for $W<W_c$ ultimately approach a universal curve (dashed black). The red dotted line is the \emph{critical} line for $W=W_c$, which separates the localized and ergodic phases. \emph{(Right)} Beta function in the absence of a localized phase. For any finite value of $W$, the RG flow never reaches $A=0$ and flows to the ergodic fixed point at large enough sizes.}
    \label{fig:beta_sketch}
\end{figure*}

\begin{figure*}
    \centering
    \includegraphics[width=0.45\textwidth]{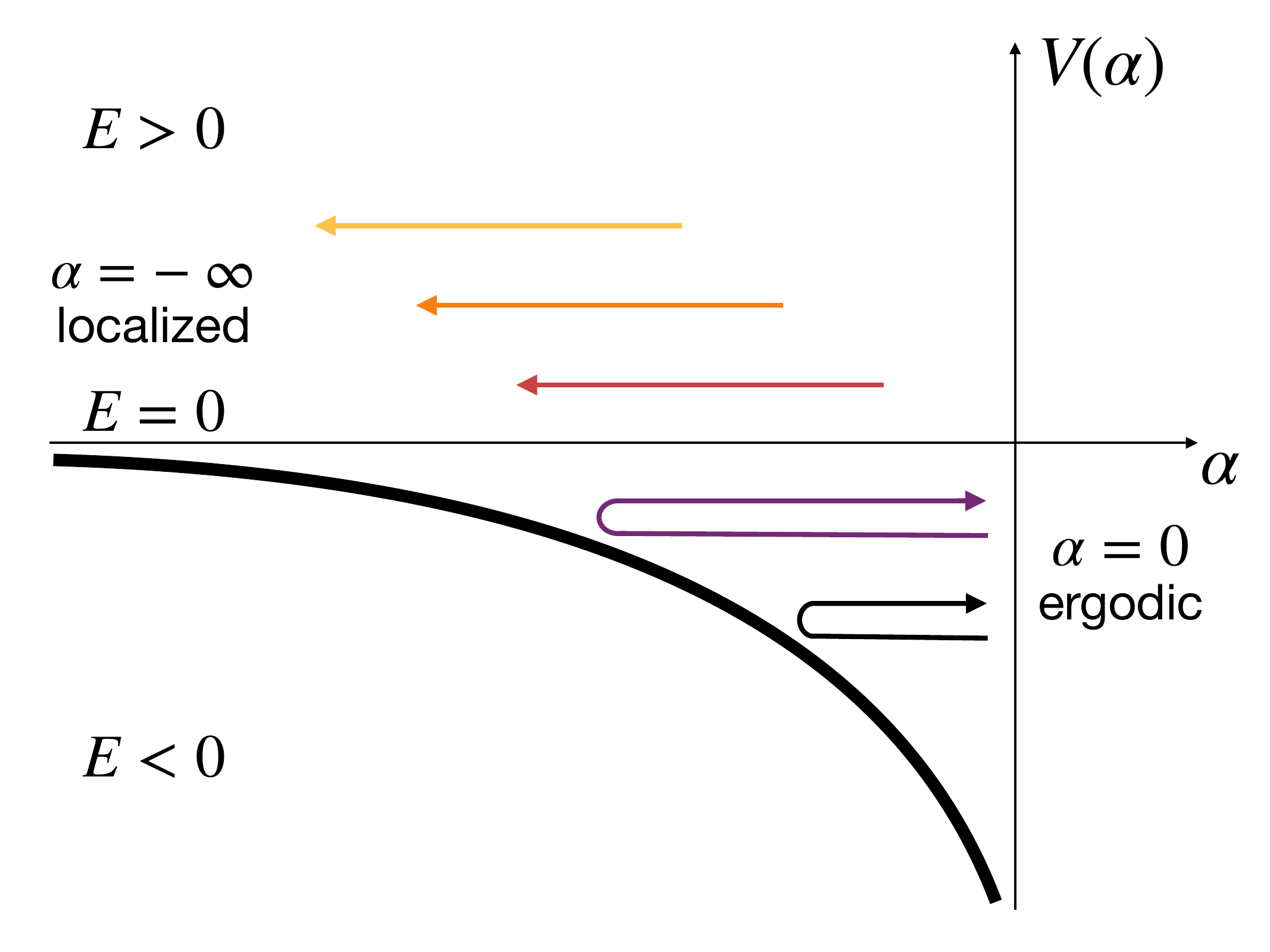}
    \includegraphics[width=0.45\textwidth]{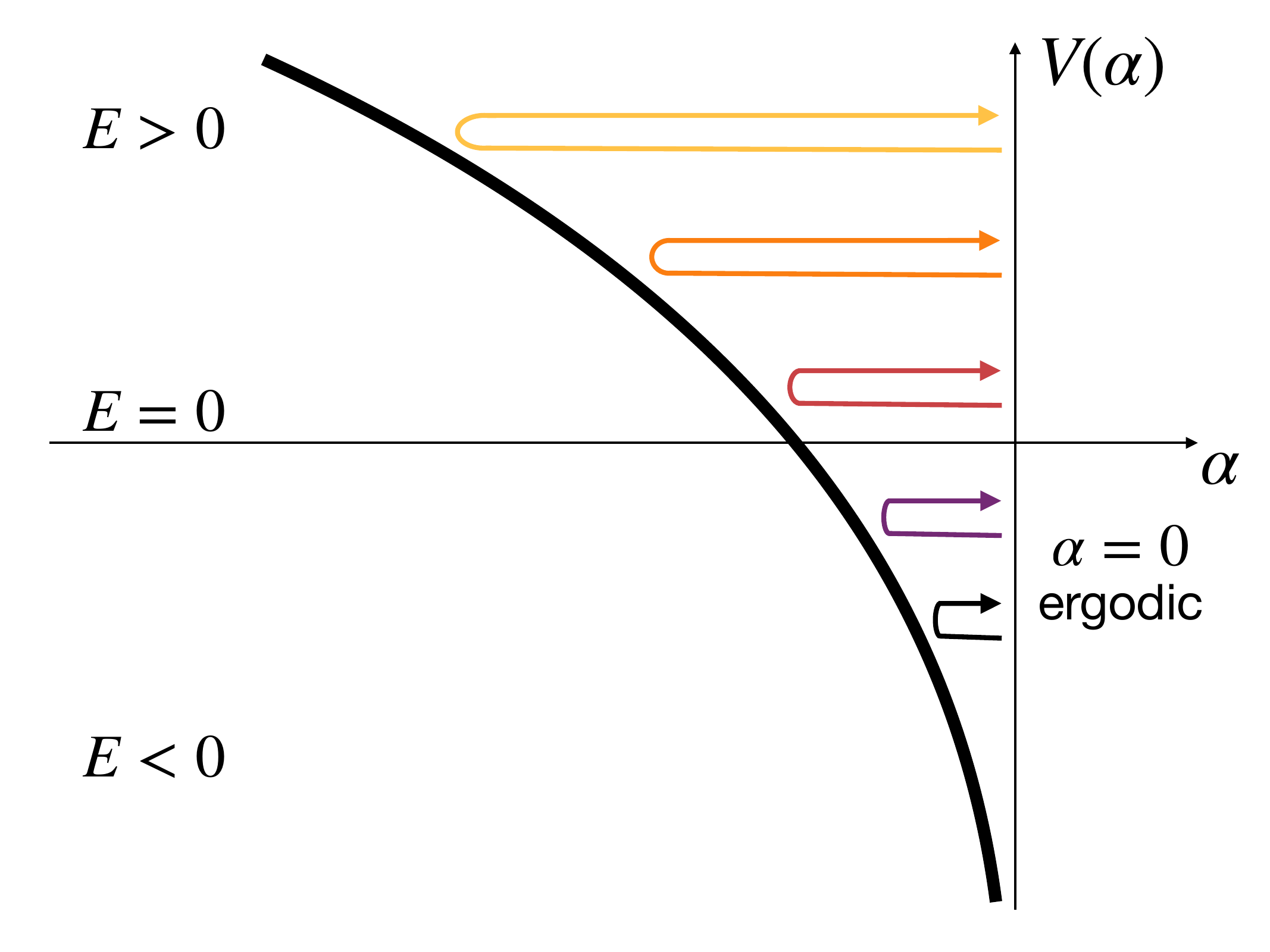}
    \caption{\emph{(Left)} Example of potential function for the one-dimensional dynamical model in presence of a localization transition, similarly to what happens for the RRG. For small disorder, corresponding to energies $E<E_c$ ($E_c=0$ in the sketch), the presence of the potential confines the trajectories, excluding the point $\alpha=-\infty$ (i.e.\ $A=0$): the original system is ergodic. For large enough disorder, corresponding to energies $E>E_c$, the potential does not confine the classical trajectories, and $\alpha \to -\infty$ when $t\to \infty$: the localized phase is approached in the thermodynamic limit. Notice also that, at $t=0$, particles move from right to left, since $\beta<0$, being the beta function negative at small enough sizes.
    \emph{(Right)} When the potential is such that $V(\alpha \to -\infty)\to + \infty$, the classical trajectories are always reflected and $\alpha \to 0$ (equivalently $A \rightarrow 1$) when $t \to \infty$. In this case there is no localized phase.}
    \label{fig:potential_sketch}
\end{figure*} 

In the previous sections, we considered the situation where there is only one relevant parameter in the Hamiltonian, and the lowest anomalous dimension is sufficiently large and negative that it can be included as a small correction to the one-parameter formulas. This occurs when the unstable fixed point $A_c$ is well separated from the two other ``Gaussian'' fixed points, which in this case represent the localized and ergodic phases. When instead $A_c\to A_\mathrm{loc}$ (like the Anderson model in $d\to\infty$) or $A_c \to A_\mathrm{erg}$ (like the Anderson model as $d\to 2$), one of the parameters must necessarily become a flat direction or, in the language of RG, {\it marginal}. As a concrete example, this happens when the Wilson-Fisher critical point merges with the Gaussian fixed point in the Ising model at the upper critical dimension~\cite{Cardy_1996}.

In order to go beyond one parameter scaling, let us define the RG time $t \equiv \ln N$ and
\begin{equation}
    \alpha \equiv \ln A.   
\end{equation}
Let us also represent with a dot the derivatives wrt.\ RG time: $d\alpha/dt \equiv \dot \alpha$. The RG equations acquire the general form
\begin{subequations}
    \begin{align}
        \dot{\alpha} &= \beta \\
        \dot{\beta} &= \gamma(\alpha, \beta)
    \end{align}
\end{subequations}
for a certain function $\gamma$. This is the general system of equations describing a two-parameter scaling RG flow. In principle, the function $\gamma$ depends also on $\beta$: this dependence is of particular importance also to model the adherence to the one-parameter scaling curve towards the ergodic fixed point as the overdamped limit of a dynamical system. However, in the region of $A$ near zero it can be seen from the numerical data that $\gamma$ is a function of $\alpha$ alone; therefore, one can reduce to the system of equations
\begin{subequations} \label{eq:Hamilton}
    \begin{align}
        \dot{\alpha} &= \beta \\
        \dot{\beta} &= \gamma(\alpha).
    \end{align}
\end{subequations}
The simplification $\gamma(\alpha,\beta) \to \gamma(\alpha)$ is crucial, because now the RG equations can be interpreted as Hamilton's equations of motion for an effective one-dimensional particle, living on the half-line $\alpha\leq 0$. Its Newton's equation reads: $\ddot\alpha = \gamma(\alpha)$, so $\gamma(\alpha)$ is the force field. RG equations that acquire Hamilton's (or Newton's) form are special because they admit an integral of motion, the energy:
\begin{equation}
    \label{eq:energy}
    E = \frac{1}{2}\dot{\alpha}^2 + V(\alpha),
\end{equation}
where $V(\alpha) = -\int \gamma(\alpha) d\alpha$. The conservation of energy is a manifestation of the time/scale invariance which arises near the critical point $A=0$. 

In order to fix ideas, it is instructive to consider the illustrative example of the Anderson model on random regular graphs (RRG)~\cite{vanoni2023renormalization}. There, the numerics for the fractal dimension is well described by $\gamma(\alpha) \propto e^\alpha$ in the region $\alpha \to -\infty$, giving rise to Liouville equation
\begin{equation}
    \label{eq:Newton_exponential}
    \ddot \alpha=c e^\alpha
\end{equation}
for some constant $c$. The energy, in turn, reads
\begin{equation}
    \label{eq:energy_exponential}
    E = \frac{1}{2}\dot\alpha^2-c e^\alpha.
\end{equation}
The orbits of this dynamical system depend strongly on the value of $E$, which is fixed by the initial microscopic parameters. If $E>0$, the particle can go all the way to $\alpha\to-\infty$, i.e.\ $A=A_\mathrm{loc}=0$; for $E<0$, it bounces back with $\alpha\to 0$, i.e.\ $A\to A_\mathrm{erg}= 1$. Therefore, one can see that the presence of two stable phases, separated by a critical point $E=0$, corresponds to whether the force field is \emph{confining} or not: confining potentials correspond to the absence of localization, while non-confining ones admit a localized phase. This is shown by the sketches in Figs.~\ref{fig:beta_sketch} and \ref{fig:potential_sketch}.

In the case of the RRG --- Eqs.~\eqref{eq:Newton_exponential} and~\eqref{eq:energy_exponential} --- the critical trajectory is $A_\mathrm{crit}(t) \propto 1/t^2$, which describes well the numerical findings~\cite{vanoni2023renormalization}. Notice that the existence of a localized region is universal for all $n>0$ in $\gamma(\alpha)=e^{n\alpha}$. In fact, one can reabsorb $n$ by rescaling $\alpha \to \alpha/n$, but this corresponds to $A\to A^{1/n}$ and therefore the critical line becomes $A_\mathrm{crit}(t)\sim 1/t^{2/n}$. The value of the exponent $n$ determines the power low behavior of observables near the critical point and thus ``labels'' different universality classes.

This example also clarifies the meaning of the second equation for $\dot \beta$: it describes the renormalization of the localization length $\xi_\mathrm{loc}$, as a function of the distance from the localized critical point. This can be seen from the fact that observables (e.g.\ the $r$-parameter) decay exponentially with the system size $L = \ln N$, where in this case $N$ is the number of vertices in the RRG. Assuming $A\propto e^{-L/\xi_\mathrm{loc}}$, one sees that $\beta = -1/\xi_\mathrm{loc}$ and $\dot\beta=\dot\xi_\mathrm{loc}/\xi_\mathrm{loc}^2$. The effects encoded in this equation are \emph{non-perturbative}, since the usual perturbation theory in the hopping term (locator expansion) only predicts exponential decay, without renormalization of the decay length. 

The example considered so far admitted the presence of a localized phase. A second, natural class of forces to consider is the one in which all trajectories go to the ergodic fixed point $\alpha=0$ (i.e.\ $A=1$), corresponding to confining potentials. A paradigmatic form which has such a behaviour is the logaritmic potential $V(\alpha)=\ln|\alpha|$, leading to Newton's equation
\begin{equation}
    \ddot\alpha=-1/\alpha.
\end{equation}
This equation can be solved by quadratures using the formula
\begin{equation}
    \int_{\alpha_0}^{\alpha(t)}\frac{d\alpha}{\sqrt{2(E-\ln|\alpha|)}}=t-t_0,
\end{equation}
since the energy reads
\begin{equation}
    E = \frac{1}{2}\dot\alpha^2+\ln|\alpha|.
\end{equation}
All the trajectories have a turning point, located at $\alpha_A=-e^{E}$, and are therefore confined as claimed above. For such potential, the localized phase---in the sense of observables exponentially decreasing with system size---does not exist. The only thing happening as the disorder is increased is a slower and slower approach to ergodicity. From the perspective of the renormalization of the localization length, the logarithmic potential entails
\begin{equation}
    \dot\xi_\mathrm{loc}\propto 1/\ln(1/A).
\end{equation}
In the limit $A\to 0$ one sees that $\dot\xi_\mathrm{loc}=0$, but such a slow convergence towards this value makes the entire localized phase {\it unstable} towards developing ergodicity.

\subsection{Two possible scenarios for MBL}

Let us here spend a few words to summarize the main results discussed in the previous sections. We have presented two possible scaling theories for localization transitions: the first one is compatible with the \emph{one-parameter} scaling hypothesis, around both the ergodic and the localized fixed points and rules the critical behaviour of the Anderson model in finite dimensions $d>2$; the second one is instead characterized by one-parameter scaling only around the ergodic fixed point, while trajectories approach a line of localized fixed points according to a \emph{two-parameter} scaling hypothesis. This is the case of the Anderson model in the limit $d\rightarrow \infty$, which is also the model defined on the RRG.

It is natural to ask what happens in many-body problems, e.g.\ the XXZ quantum spin chain. In this case, we can draw two possible scenarios, which will be addressed by our numerical analysis in the following section. The first is the \emph{transition scenario}: if a localization transition occurs, the critical point has to be localized (as we will also discuss in the following) and, consequently, a two-parameter scaling theory describes the critical properties. 
This case corresponds to the beta function cartoon on the left in Fig.~\ref{fig:beta_sketch}: the ergodic trajectories are separated from the localized ones by a critical line which terminates on the point $(A=0,\beta=-1/\xi_\mathrm{loc}=0)$, which is the last of a line of localized critical points $(A=0,\beta=-1/\xi_\mathrm{loc}<0)$. This scenario is well captured by a non-confining potential as the one sketched in the left panel of Fig.~\ref{fig:potential_sketch}. The second scenario, on the other hand, is the \emph{no-transition scenario}: in this case all the trajectories, despite displaying finite size localization, will flow eventually to the ergodic fixed point $A=1$, governed by a one parameter curve in the large size limit (see right panel of Fig.~\ref{fig:beta_sketch}). The no-transition scenario is well captured by a confining potential as the one sketched on the right in Fig.~\ref{fig:potential_sketch}.  

In the following Sec.~\ref{sec:numerical_results} we present numerical data that will be analyzed in the light of the observations made above.

\section{Numerical results}
\label{sec:numerical_results}

\begin{figure}
    \centering
    \includegraphics[width=\columnwidth]{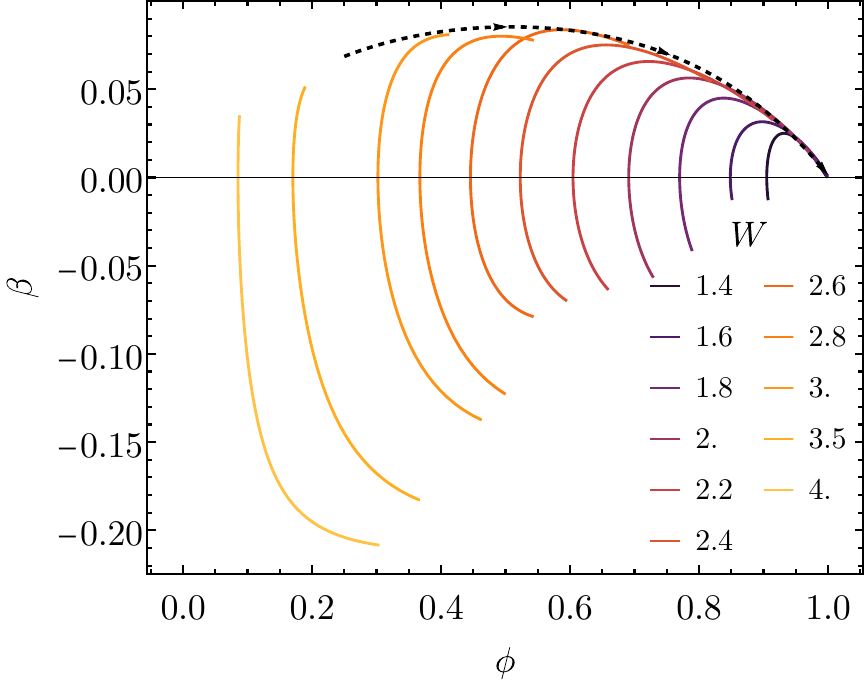}
    \caption{Beta function of the rescaled $r$-parameter in the ergodic regime. Each line represents the flow of the observable towards the ergodic fixed point, and corresponds to a different value of the disorder strength $W$. The lines are parametrized by the system size (sizes increase from bottom to top). The figure is obtained by plotting the logarithmic derivatives in $N$ of the fitting functions used in Fig.~\ref{fig:XXZ-rpar}, according to the definition Eq.~\eqref{eq:1p_scaling}. The dashed black line is the envelope of the curves obtained numerically, and represents our best estimate of the one-parameter scaling curve $\beta_0(\phi)$.}
    \label{fig:XXZ-rpar-betafunc}
\end{figure}

We now analyze the beta function for $\phi$, the rescaled $r$-parameter defined in Eqs.~\eqref{eq:r_L_def}--\eqref{eq:def_phi}. Two regions can be identified in the RG flow, which will be discussed separately: in the (finite-size) ergodic region, the RG trajectories flow toward the fixed point at $\phi = 1$ reaching an asymptotic limit described by a one-parameter-scaling curve; in the (finite-size) localized region (where $\phi \ll 1$), the RG trajectories follow a two-parameter-scaling scenario, and flow to a line of fixed points terminating at the critical point $(\phi = 0, \beta = 0)$.  The beta function is shown in Fig.~\ref{fig:XXZ-rpar-betafunc} for the ergodic regime, where it is computed using the definition~\eqref{eq:1p_scaling} \footnote{The logarithmic derivatives in Eq.~~\eqref{eq:1p_scaling} are computed by using $N=2^L$ in place of $N = \binom{L}{L/2}$, as they differ by a proportionality constant.}. We refer to Fig.~\ref{fig:beta_orbits_W} for the beta function in the localized region, to compute which we resorted to a more refined technique, discussed in Sec.~\ref{sec:largeW}.

\subsection{The ergodic fixed point}
\label{sec:ergodicFP}

From the numerical data in Fig.~\ref{fig:XXZ-rpar-betafunc}, one can observe that in the XXZ spin chain the RG curves flow towards the ergodic fixed point by eventually attaching to a one-parameter-scaling curve $\beta_0(\phi)$. As the disorder $W$ increases, the RG time needed to reach $\beta_0(\phi)$ also increases.

The one-parameter-scaling component $\beta_0(\phi)$ of the beta function can be obtained from the data by considering the envelope of the curves towards the ergodic fixed point, leading to the black dashed curve in Fig.~\ref{fig:XXZ-rpar-betafunc}. In practice, it is obtained by binning the values of $\phi$ and considering the maximum among the data corresponding to different $W$ in each bin. The list of maxima is then interpolated with a polynomial function with simple zeros in $\phi=0$ and $\phi=1$:
\begin{equation}
    \label{eq:beta0_fit}
    \begin{aligned}
        \beta_0(\phi) &= a_1 \phi (1-\phi)+ a_2 \phi (1-\phi)^2 \\
        &\hspace{0.5cm} + a_3 \phi (1-\phi)^3 + a_4 \phi^2 (1-\phi) \\
        &\hspace{0.5cm} + a_5 \phi^2 (1-\phi)^2 + a_6 \phi^3 (1-\phi).
    \end{aligned}
\end{equation}
This is the best lower bound to the one-parameter scaling curve $\beta_0(\phi)$ that can be obtained with our data: the curves at $W=3.5$ and $4$, which reasonably also flow to $\phi=1$ for system sizes larger than the ones numerically accessible, may produce an envelope which is greater than the dashed curve in Fig.~\ref{fig:XXZ-rpar-betafunc}, especially far from the region $\phi \simeq 1$.

Denoting with $N_0$ the smallest value of $N$ allowing to approximate the flow with the one-parameter flow, i.e.\ $\beta(\phi,L>L_0) \simeq \beta_0(\phi)$, one has
\begin{equation}
    \frac{d\ln \phi}{d\ln N}=\beta_0(\phi),
\end{equation}
from which 
\begin{equation} \label{eq:scaling_f}
    \phi(W,L)=f(N/N_0e^{G(\phi_0)}),
\end{equation}
where $f(x)=G^{-1}(\ln(x))$, $G(x)=\int^x d\phi/\phi\beta_0(\phi)$ and $G^{-1}$ is the inverse function. Notice that the scaling form is obtained by dividing the volume $N$ by a critical volume $N_0$, which means
\begin{equation}
    N/N_0\simeq 2^{L-L_0},
\end{equation}
which needs to be contrasted with $L/L_0$. 

\begin{figure}
\centering
    \includegraphics[width=0.98\columnwidth]{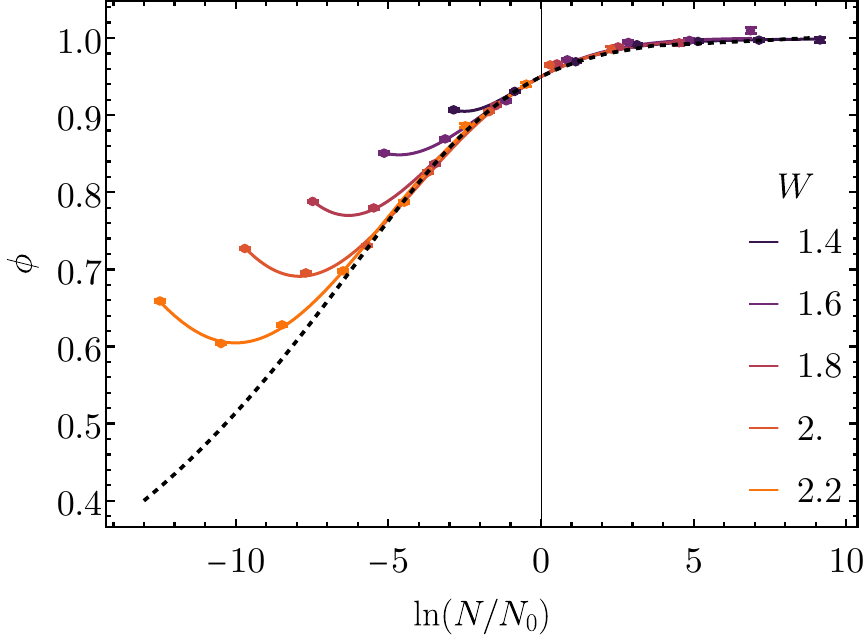}
    \caption{Data collapse near the ergodic fixed point. The black dashed curve is the scaling function $f$ obtained by integrating $\beta_0$ from Eq.~\eqref{eq:beta0_fit}. Data are collapsed according to Eq.~\eqref{eq:scaling_f} and as explained in the main text.}
    \label{fig:coll_ergod}
\end{figure}

The collapse near the ergodic fixed point is shown in Fig.~\ref{fig:coll_ergod}. For each $W$, we determine $\ln N_0(W) = L_0(W)$ such that $\phi(W,L_0(W)) = 0.95$, so that all the curves are at the same point on the one-parameter arc $\beta_0(\phi)$. The specific value chosen for $\phi(W,L_0(W))$ is not important, provided that it is large enough to allow all curves to be in the one-parameter scaling regime.

Far from the ergodic fixed point $\phi=1$, the functions $\phi(W,L)$ are regular and non-vanishing, and therefore the divergent lengthscale is given by $N_0$. The divergence of $N_0$ is entirely due to the two-parameter flow which goes close to $\phi=0$ and therefore needs to be determined by this.

\begin{figure}
    \centering
    \includegraphics[width=0.98\columnwidth]{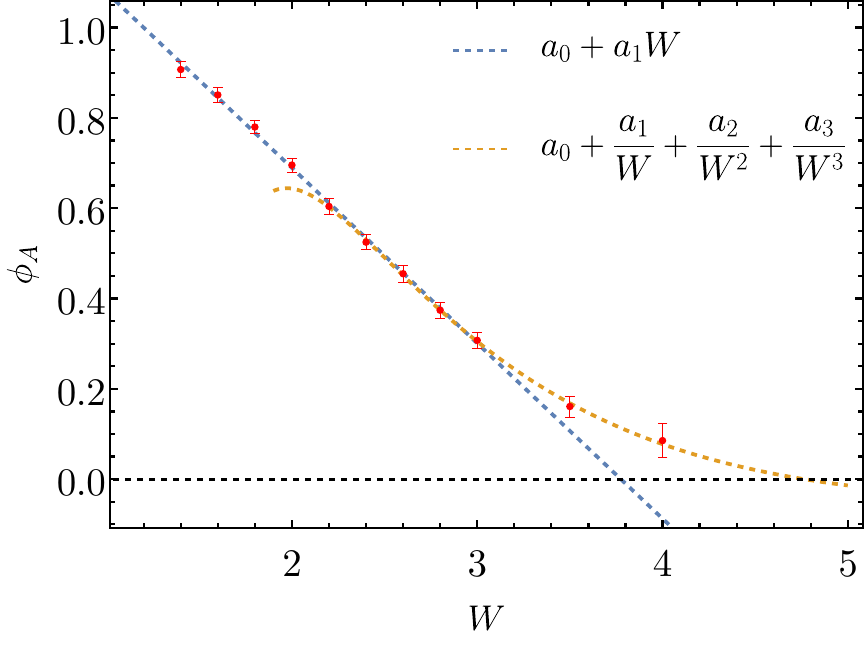}
    \caption{Values of $\phi$ at which the beta function vanishes, i.e.\ $\beta(\phi_A)=0$, as a function of the disorder $W$. The linear fit (blue) is obtained by considering points up to $W=3$ and yields a critical value of the disorder $W_c \simeq 3.8$, which is consistent with the old estimate of Ref.~\cite{Luitz15}. The nonlinear fit (yellow) is obtained by fitting points with $W \geq 2.2$ and extrapolates to a larger value of the critical disorder $W_c \simeq 4.7$.}
    \label{fig:phiA_vs_W}
\end{figure}

It is also instructive to plot, as a function of the disorder strength $W$, the values $\phi_A$ at which the beta function vanishes, i.e.\ $\beta(\phi_A)=0$. If there was a transition, one should observe a zero of the function $\phi_A(W)$: extrapolating the data to the thermodynamic limit, one would then be able to pinpoint a localization transition. The points $\phi_A(W)$ displayed in Fig.~\ref{fig:phiA_vs_W} clearly display a curvature, not crossing the axis $\phi=0$ at a finite value of $W$. It is thus not possible to use a simple linear fit to predict the critical disorder $W_c$, and different fitting functions lead to different values of $W_c$, possibly also infinite. The same analysis, when performed for the Anderson model on RRGs, shows instead no sign of curvature, and from a simple linear extrapolation, one is able to extract the value of $W_c$~\cite{vanoni2023renormalization}. Such value agrees with the critical disorder strength predicted by the analytical computation on the Bethe lattice~\cite{abou1973selfconsistent,parisi2019anderson} with an error of about $1\%$. Unfortunately, for the XXZ chain, there is no independent method of extracting the critical disorder value, and one is not able to assess the presence of a transition by looking only at the small disorder data. In the next section, we complement the analysis by looking at larger disorder strengths, and with the help of the two-parameter-scaling formalism, we analyze the possibility of a finite-$W_c$ localization transition.

\subsection{Large disorder: Localized and critical behavior} \label{sec:largeW}

The numerical analysis of the $\beta$ function in the critical localized region---i.e.\ the region defined by $\beta<0$ and $\phi \ll 1$---presents an intrinsic difficulty, since the system sizes required are too large to be accessed numerically. In the following, we show what the expected behavior is for an observable near the critical region, given the dynamical system formulation developed in Sec.~\ref{sec:1to2PS}. 

Upon defining $\alpha=\ln A=\ln\phi$, the two RG equations can be put in the form Eq.~\eqref{eq:Hamilton}. The force $\gamma(\alpha)=-V'(\alpha)$ is to be determined from the numerical data. As discussed before, a confining potential corresponds to the absence of a transition, while a non-confining potential corresponds to the existence of a transition.

\begin{figure}
\bigskip
    \centering
    \includegraphics[width=0.9\columnwidth]{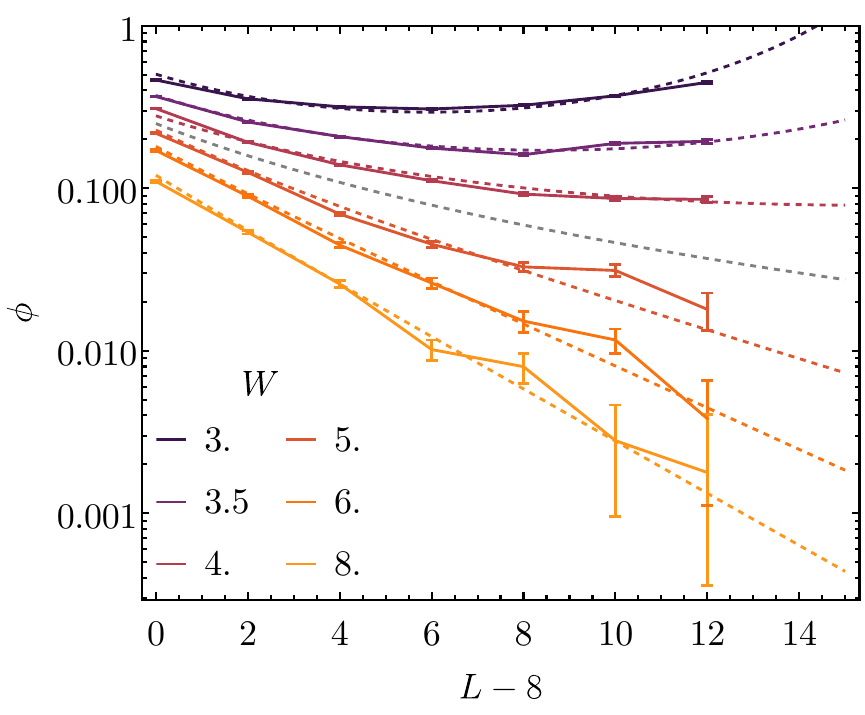}
    \caption{Comparison between numerical results and trajectories obtained through the integration of the dynamical system, Eq.~\eqref{eq:Hamilton}, with potential function $V(\alpha) = c\, e^{n \alpha}/n$, where $\alpha=\log \phi$. All the orbits are consistent with a scaling coefficient $c \simeq 0.1$ and an exponent $n \simeq 1$. The dashed gray line is the critical line according to Eq.~\eqref{eq:crit_line}, with $\phi_0 \simeq 0.25$.}
    \label{fig:dynsyst_W}
\end{figure}

In order to explore the possible behaviors of $\alpha$, we start by solving Hamilton's equations~\eqref{eq:Hamilton} with a force $\gamma(\alpha)=c e^{n\alpha}$. Later on, we will comment on the choice of this functional form. The exponent $n$ and the coefficient $c$ are left free for now, and will be fixed by confronting to the numerical data. This exponential force $\gamma$ comes from a potential $V(\alpha)=-(c/n)e^{n\alpha}$, and therefore the conservation of energy can be written as
\begin{equation}
    \label{eq:energy_pot_phi^n}
    \frac{\beta^2}{2} - \frac{c}{n} \left(\phi^n + \frac{n}{c}E \right) = 0.
\end{equation}
From this, one finds that the beta function is given by 
\begin{equation}
    \label{beta_pot_phi^n}
    \beta(\phi) = \pm \sqrt{\frac{2c}{n}} \sqrt{\phi^n + \frac{n}{c}E} \; ,
\end{equation}
where negative $E$ corresponds to the ergodic phase (which possesses both branches $\beta>0$ and $\beta<0$), while positive $E$ corresponds to the localized phase (for which $\beta<0$). The region $\phi\lesssim |\frac{n}{c}E|^{1/n}$ corresponds to the critical region.

\begin{figure}
\bigskip
    \centering
    \includegraphics[width=0.9\columnwidth]{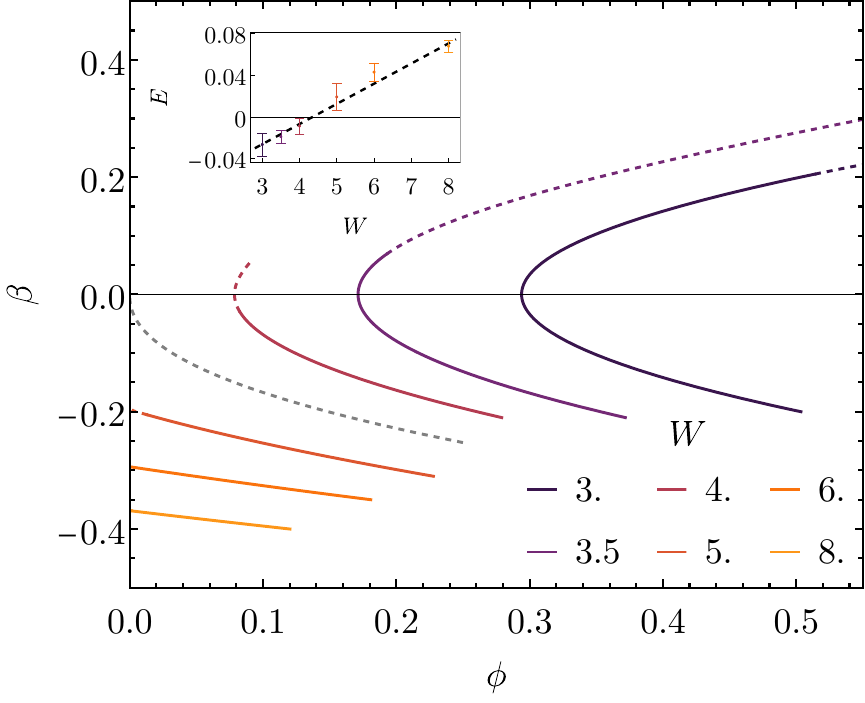}
    \caption{Beta function of the rescaled r-parameter for the disorder strengths $W$ in legend, obtained as a phase-space plot of the trajectories in Fig.~\ref{fig:dynsyst_W}. Lines are continuous up to the point where data are available. The dashed gray line is the critical line, separating the ergodic from the localized behavior. In the inset, the energies of each trajectory are shown as a function of the disorder, with the same color map. The linear fit yields a value $W_c = 4.4(4)$.}
    \label{fig:beta_orbits_W}
\end{figure}

Let us focus on the critical curve $E=0$. If one integrates Eq.~\eqref{beta_pot_phi^n}, using the definition $\beta=\dot{\phi}/\phi$, one finds an algebraic scaling with the system size
\begin{equation}
    \beta = - \sqrt{\frac{2c}{n} }\phi^{n/2},
\end{equation}
and in turn
\begin{equation}
\label{eq:crit_line}
    \phi(L)=\frac{\phi_0}{(1+\sqrt{\frac{cn}{2}}\phi_0^{n/2}(L-L_0))^{2/n}}, 
\end{equation}
where $(\phi_0,L_0)$ are the initial conditions ($\phi_0$ is a microscopic function of $W,L_0$ which can be obtained from the numerics). For large $L\gg L_0$, one finds a power-law decay
\begin{equation}
\label{eq:critical_phi}
    \phi(L)\sim L^{-2/n}.
\end{equation}
This is the same scenario found for the Anderson model on the RRG, where the data support $n=1$ and a law $\phi\sim 1/L^2$~\cite{vanoni2023renormalization}. We will see in a moment that $n=1$ is also compatible with the numerical data for the XXZ chain.

Inside the localized region, it holds $E>0$. If the starting point has $\phi\gg (ncE/2)^{2/n}$, there is an initial decay which follows the critical law Eq.~\eqref{eq:critical_phi}, after which an exponential scaling sets:
\begin{equation}
  \beta \simeq - \sqrt{\frac{2c}{n}} |\phi_A|^{n/2} \equiv -1/\xi_{\mathrm{loc}} \; , \quad \phi(L) \sim e^{-L/\xi_\mathrm{loc}} \; .
\end{equation}

In Fig.~\ref{fig:dynsyst_W} we show a comparison between the trajectories obtained from the integration of the equations of motion and the numerical data. We have also included data corresponding to intermediate values of the disorder, to check whether it is possible to fit them with the same potential function. The free parameters $n,c$ are obtained from a best fit of the data corresponding to small values of $\phi$, or moderately large disorder values $W=3.5,...,8$. The best fit values are $n= 1.0 (1),\ c=0.10 (5)$. 

Once $n,c$ are fixed, the best fitting curves are obtained by choosing initial conditions $(\phi_0,\beta_0)$ for each $W$. In this way, a value of the energy can be associated with each trajectory by using Eq.~\eqref{eq:energy_pot_phi^n} at the initial condition: we find $E<0$ for disorder values $W \leq 4$ and $E>0$ for $W \geq 5$. This places the value of the critical disorder between $W=4$ and $W=5$. The critical line Eq.~\eqref{eq:crit_line} is drawn by fixing $\phi_0$ to a value such that that line does not cross the other trajectories: this results in a tiny interval of possible values around $\phi_0 \simeq 0.25$. 

In Fig.~\ref{fig:beta_orbits_W} we show the phase space plot $(\phi,\beta)$ constructed from Fig.~\ref{fig:dynsyst_W}. Notice that the curves possessing a positive energy (corresponding to $W=5,6,8$) attach to the negative $\beta$ axis. On the other hand, the curves possessing a negative energy ($W=3,3.5,4$) are confined by the potential and, though starting with a negative $\beta$, reach a turning point and flow away from the localized phase. The critical line Eq.~\eqref{eq:crit_line} yields the separatrix between the ergodic and the localized behaviour.

Let us stress that, although somewhat arbitrary, the choice of the potential function family $V(\alpha)=-(c/n)e^{n\alpha}$ to fit the data is the simplest option that is, \textit{a priori}, compatible with both the transition and no-transition scenarios, depending on whether $n$ is positive or negative. This choice is motivated by physical intuition, namely the approximate similarity between the topology of the Hilbert space of the Anderson model in infinite dimension and that of the Fock space of the XXZ chain \cite{altshuler1997quasiparticle,de2014anderson}. The outcome of the fitting procedure corroborates this intuition. While more elaborate functional forms might yield slightly better numerical fits, such fits would be entirely phenomenological and would lack the suggestive physical interpretation provided by our approach.

\subsection{Significance of the results at large disorder: a problem of statistics}

Resolving the value of $\phi$ for large disorder necessitates a heavy numerical effort. In fact, since the critical point is the final point of a line of fixed points, one needs to distinguish between two different {\it decays} of $\phi(L)$, one in the critical region $W\simeq W_c$ and the other in the MBL phase $W\gg W_c$. If this is not done, one cannot {\it bona fide} claim that the transition is not just a crossover. This brings up the necessity to have very precise values of $\phi$ (or any other order parameter), whose value in the MBL region is zero.

After an initial power-law decay $\phi\sim L^{-\gamma}$ (with $\gamma=2/n$, the critical power law obtained from (\ref{eq:Hamilton})) the system either reverses the course and flows back to $\phi=1$ at the ergodic fixed point, or settles on an exponential decay that will follow asymptotically for $L\to\infty$. At what values of $\phi,W,L$ this occurs is dictated by the value of $W_c$. If this is too large, $\phi$ will be very small already at small $L$ (say $L\leq 14$), and a huge sample is needed to distinguish the signal from the noise. In this section, the Kolmogorov-Smirnov hypothesis test is used to estimate how large a sample one needs to distinguish the data from a Poisson distribution, and therefore the significance of the value of $\phi$ itself.

The probability distribution function of the $r$-parameter for an integrable system is (see e.g.\ Ref.~\cite{Bogomolny2011Integrable})
\begin{equation}
    p(r) = \frac{2}{(1+r)^2} \; , \qquad r \in [0,1] \;.
\end{equation}
This law is very broad in its domain; hence, the average value might not be an efficient numerical estimator. It is then convenient to quantify the difference between the numerical and the theoretical probability distribution function beyond the average value. This way, it is possible to provide an estimate of how much statistics is needed to have a significant dataset. To do so, we compute the difference between the theoretical and numerical cumulative distribution functions:
\begin{equation}
    \kappa(r) \equiv \left| C^\mathrm{P}(r) - C^\mathrm{exp}(r) \right| \; ,
\end{equation}
where $C^\mathrm{P}(r)= 2r/(1+r)$ and $C^\mathrm{exp} = \{ \sum_i n_i < r \}/ N$. This choice is motivated by a physical argument: it is expected that, for small system sizes and in the large disorder limit, the system still exhibits some level repulsion, which is reflected in the small values of $r$. The error associated to $C^\mathrm{exp}$ is a simple Poissonian counting error: $\delta C^\mathrm{exp}(r) = \sqrt{C^\mathrm{exp} / N} $.

\begin{figure}
 \bigskip
    \centering
    \includegraphics[width=1\linewidth]{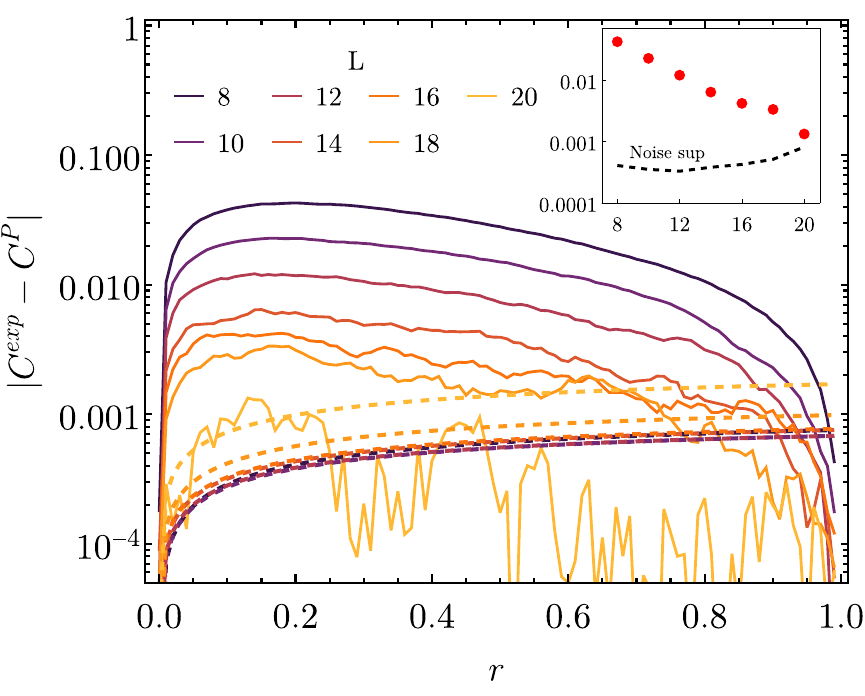}
\bigskip
    \includegraphics[width=1\linewidth]{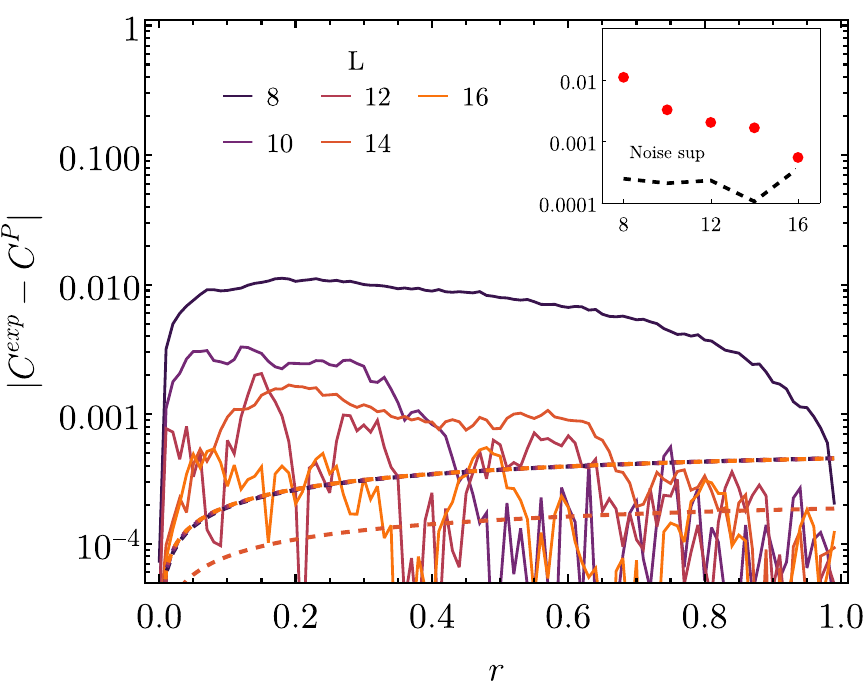}
        \caption{Comparison of signal and noise for $W=6$ (top) and $W=20$ (bottom). The solid lines represent the numerical data, while dashed lines represent the Poissonian error associated with the signal (with matching color code). In the insets, the behavior of $\kappa_s(L)$ is compared to its associated error. }
    \label{fig:significance 20.0} 
\end{figure}

In Fig.~\ref{fig:significance 20.0} we compare the signal and noise from our data as a function of the system size, for the particular cases of $W=6$ and $20$. In the case of extremely large disorder $W=20$, we considered a larger number of samples (from $10^5$ to $6 \times 10^5$) to show that the difficulty of collecting meaningful statistics increases both with the disorder and with the system size.

The condition to have a significant dataset is that the signal is much greater than the noise, i.e.\ $ \kappa(r) \gg \delta C^\mathrm{exp}(r)$. One can be more quantitative considering $\kappa_s = \sup_{r \in [0,1]} \kappa (r)$: $\kappa_s$ is called the Kolmogorov-Smirnov parameter. 

We compute $\kappa_s(L)$ and compare it with $\delta C^\mathrm{exp}(r_s)$, where $r_s$ is the value at which $\delta C^\mathrm{exp}(r)$ is maximal. The data at $W=6.0$ clearly shows exponential decay of the Kolmogorov-Smirnov value all the way from $L=8$ to $L=20$, therefore confirming the existence of a transition at $W<6$. The data at $W=20$ instead fall quickly into the region where the signal is comparable to the error.

We can provide a rough estimate of the size of a significant dataset $N_\mathrm{sig}$ extrapolating with an exponential fit $\kappa_s(L)$ for $W=20$ (inset of the second panel of fig $\ref{fig:significance 20.0}$), and considering the point at which the signal is equal to the statistical error: $\sqrt{N_\mathrm{sig}} \sim \delta C^{-1} \sim k_s(L) $. We find $N_\mathrm{sig} \sim e^{0.6 L}$; meaning that, with $50$ eigenvalues per sample,  the number of samples needed to have an accurate estimate of the average value of the $r$-parameter at $L=20$ is $\gtrsim 10^6$. This {\it bona fide} computation shows that a meaningful statement is quite hard to make, even in the localized region.

\section{Discussion and Conclusions}
\label{sec:conclusions}

Motivated by the recent progress in the scaling theory of Anderson localization transitions \cite{altshuler2024renormalization,vanoni2023renormalization} and by the even more recent analytical work on the Imbrie model \cite{deroeck2024absence}, 
we developed the renormalization group (RG) analysis of the many-body localization (MBL) transition in the random-field XXZ spin chain. The presence of a localization transition in this model, though initially assessed by analytical (approximate) and numerical studies, has been undermined by general arguments based on non-perturbative effects, such as many-body resonances and thermal avalanches, and still remains debated in the community. With this work, we provide a novel contribution to the debate by focusing on the scaling theory and giving particular attention to the methodological aspects of analyzing numerical data for MBL at the small system sizes that are currently available with the present technology.

By exploiting numerical data for the rescaled $r$-parameter $\phi$, Eqs.~\eqref{eq:r_L_def}--\eqref{eq:def_phi}, obtained via exact diagonalization, our approach allows for the reconstruction of the beta function of the model. This is an essential tool for the understanding of the scaling theory of phase transitions in general, and localization transitions in particular. In fact, as discussed in Sec.~\ref{sec:crit_point}, the properties of the beta function determine the nature of the scaling functions with which one can obtain a good data collapse near the critical point. If one assumes that the critical point is a simple zero of the beta function, and therefore the flow to the attractive fixed points is ruled by a one-parameter scaling curve (as it usually happens in critical phenomena), then one is led to inconsistent data collapses in the case of MBL, as the one shown in Fig.~\ref{fig:scalingXXZ}.

The main message of this work can be summarized as follows. There are two possible scenarios for the fate of MBL in the XXZ spin chain: either there is no transition in the thermodynamic limit and the observed localized regime is just a pre-asymptotic effect which is eventually destabilized non-perturbatively; or, if there is a stable localized phase, then the transition is characterized by a two-parameter scaling towards a line of fixed points, of which the critical point is the endpoint, in a way that resembles the Berezinskii–Kosterlitz–Thouless (BKT) transition. This critical theory, which has also been observed in the Anderson model on random regular graphs (RRG)~\cite{vanoni2023renormalization}, is more difficult to study with respect to usual critical phenomena. Since the critical point coincides with a localized fixed point, the system is both repulsed and attracted in the critical region: to understand whether the system is actually flowing towards the localized phase means being able to distinguish between a power-law or exponential decay. This is a very subtle task, which requires having access not only to larger system sizes, but also to huge statistics for the higher accessible sizes.

We therefore analyze the data in the light of the two-parameter scaling transition scenario. We divided our analysis by focusing first on the flow to the ergodic fixed point for small disorder values $W$ in Sec.~\ref{sec:ergodicFP}, and then on the localized behaviour at larger disorder values in Sec.~\ref{sec:largeW}. The former is governed by a one-parameter scaling function $\beta_0$, to which each finite-size curve attaches for a certain value of the Hilbert space volume $N_0(L,W)$. A consistent data collapse of the values of the rescaled $r$-parameter $\phi$ near the ergodic fixed point can be obtained by exploiting the one-to-one correspondence between $\beta_0(\phi)$ and the scaling function $f$ used for the collapse. The volume $N_0$ sets a scale (i.e.\ a correlation length) which is larger and larger upon increasing $W$. The divergence of $N_0$ reflects the fact that the observable displays a minimum which is more and more flat as $W$ grows. The question is then: is there a finite value of the disorder for which the minimum shifts to the localized value at infinite $N_0$?

In order to answer this question, we fitted the numerical data for large disorder by adopting a more ``physical'' point of view, rather than just using some fitting functions depending on many parameters. In fact, our purpose has been to avoid nasty fluctuations in the data and to find their general trend, in a way that is grounded in the understanding of the scaling theory. For this purpose, we realized that the phenomenological equations of the two-parameter scaling of a generic observable $A$ can be cast into the equations of motion for a Newtonian one-dimensional particle with coordinate $\alpha=\ln \phi$. The Hamiltonian part of these equations involves a force term, which can be derived from a potential $f(\alpha)=-V'(\alpha)$, and rules the scaling of the observable near the critical point and in the localized region. Whether the potential is non-confining/confining corresponds to the presence/absence of the transition: in the first case, there is a critical energy (related to the critical disorder $W_c$) above which orbits can escape the potential and have access to the localized phase.

We focused on the Hamiltonian dynamics for intermediate and high values of the disorder strength, and found the free parameters $n$ and $c$ of the non-confining potential $V(\alpha)=-c e^{n \alpha}$ that allow for a better fit of the data. It turned out that $n \simeq 1$, leading to a scaling near the critical point of the kind $\phi \sim 1/L^2$, which is precisely the same scaling observed in the Anderson model on RRGs. This suggests that the two models may belong to the same universality class. From the sign of the energy labelling the orbits, we were able to assess that in this scenario the critical disorder should be between $W=4$ and $5$. This is confirmed by a fit of the minima of $\phi$, i.e., the zeros of the beta function at finite size. 

Given the fact that the observable is flowing to its localized value at the critical point, it is challenging to resolve an exponentially decaying observable from the statistical fluctuations in the data. For this reason, we performed a careful analysis of the signal and noise extracted from our data by means of the Kolmogorov-Smirnov test. 

Summarizing, we believe that the study of the full RG beta function is the correct way to address ergodicity-breaking transitions in quantum systems, lacking strong analytical arguments in favour of a simple one-parameter scaling fixed point. In the case of the XXZ model, our study of the beta function points quite strongly against the possibility of having such a simple critical fixed point. This observation has important consequences, namely, if a localization transition occurs at finite $W_c$, it must be described by two-parameter scaling, and we provided a suitable candidate that fits the data. If a localization transition {\it does not} exist at any $W_c$ then this is an even stronger reason to prefer a two-parameter scaling, and we have shown concretely how to describe this scenario. Further work is needed to resolve the correct two-parameter RG flow for the XXZ (our analysis suggests that this is within reach of existing computing power), and for the other many-body models showing breaking of ergodicity due to disorder (for example \cite{Laumann2014MBMobility, 
ponte2017thermal,suntajs_vidmar_22,crowley2022partial}).  

\acknowledgements

We would like to thank G.\ Parisi, S.\ Pascazio, and L.\ Vidmar for discussions and collaboration on related projects.

The numerical simulations were performed using the libraries  PETSc~\cite{PETSc} and SLEPc~\cite{SLEPc} on the SISSA cluster Ulysses and on the University of Bari and INFN cluster ReCaS \cite{ReCaS}. 

The work of JN was funded by the European Union--NextGenerationEU under the project NRRP Project ``National Quantum Science and Technology Institute" — NQSTI, Award Number: PE00000023, Concession Decree No.~1564 of 11.10.2022 adopted by the Italian Ministry of Research, CUP J97G22000390007. The work of AS was funded by the European Union--NextGenerationEU under the project NRRP ``National Centre for HPC, Big Data and Quantum Computing (HPC)'' CN00000013 (CUP D43C22001240001) [MUR Decree n.\ 341--15/03/2022] -- Cascade Call launched by SPOKE 10 POLIMI: ``CQEB'' project. GM is partially supported by the Italian funding within the ``Budget MUR - Dipartimenti di Eccellenza 2023- 2027'' (Law 232, 11 December 2016) - Quantum Sensing and Modelling for One-Health (QuaSiModO), from INFN through the projects ``QUANTUM'' and ``NPQCD'', and from the University of Bari via the 2023-UNBACLE-0244025 grant. \\

\appendix

\section{Details of the numerical procedure}
\label{app:sec:EDspec}

We performed an exact diagonalization algorithm using the shift-invert method, implemented on the libraries SLEPc~\cite{SLEPc}, PETSc~\cite{PETSc}. The main numerical bottleneck is the amount of memory needed to solve the linear system to invert the resolvent operator \cite{Pietracaprina2018Shift}, that scales as $\sim \mathrm{dim}(\mathcal{H})^3 $. In the present case, we focused on the spin-0 sector, which is the largest symmetry sector for even sizes, for which we remind that $\mathrm{dim}(\mathcal{H}) = \binom{L}{L/2}$. 

In the following table we present a summary of the data that we used to perform the analysis.\\

\begin{table}[h]
    \centering

    \begin{tabular}{@{}ccc@{}}
        \toprule
        \textbf{Size} & $\quad$ \textbf{No. of samples} $\quad$ & \textbf{RAM required} \\ \midrule
        $8$     &  $5\times10^4$  & $\mathcal{O}(1) \;\mathrm{MB}$             \\
        $10$          &  $5\times10^4$  & $\mathcal{O}(10)\;\mathrm{MB}$                    \\
        $12$           &  $5\times10^4$     & $\mathcal{O}(10) \;\mathrm{MB}$           \\
        $14$    &  $5\times10^4$   &$\mathcal{O}(100) \;\mathrm{MB}$                   \\ 
        $16$        &  $3\times10^4$ & $\mathcal{O}(100) \;\mathrm{MB}$ 
                \\
        $18$          &  $2\times10^4$ & $ 2 \;\mathrm{GB}$                   \\ 
        $20$          &  $1.5\times10^4$  & $ 30 \;\mathrm{GB}$                  \\ 
        
        \bottomrule
    \end{tabular}
    \caption{Approximate number of samples for each size $L$ and value of disorder $W$, and RAM required for the diagonalization of a single sample.}
    \label{tab:EDspecs}
\end{table}

\section{Data collapse in the one-parameter-scaling framework}
\label{app:sec:datacollapse}
\begin{figure}
 \bigskip
    \centering
    \includegraphics[width=1\linewidth]{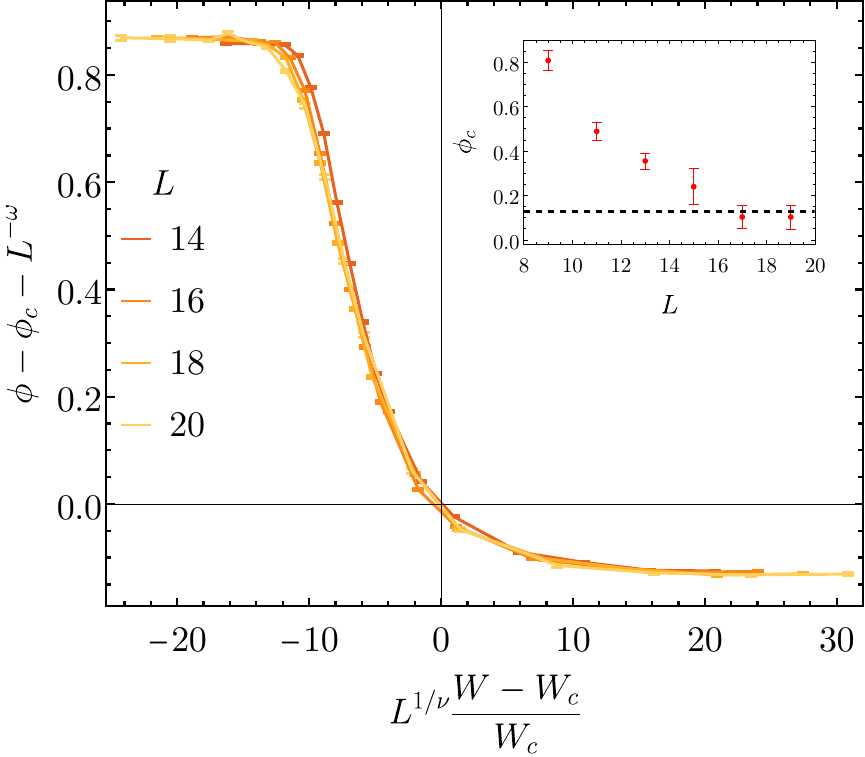}
    \caption{Data collapse near the critical point according to the one-parameter scaling hypothesis of the first works on MBL \cite{Deluca13,Luitz15}, estimating $W_c \simeq 3.8$, $\nu \simeq 0.9$, and $\omega \simeq 2$. The dashed line denotes the ``fake'' critical value $\phi_c \simeq 0.13$. The two-parameter scaling discussed in this paper starts from the assumption that $\phi_c=0$. In the inset the drift of the crossing point between the rescaled $r$-parameter of adjacent sizes is shown. The dashed line in in $\phi_c \simeq 0.13$, which is compatible with the error bars of the the last two crossing points.}
    \label{fig:r_par_collapse-Luitz} 
\end{figure}

In this appendix we show the other data collapse mentioned in the main text in Sec.~\ref{sec:crit_point}, which wrongly assumes a one parameter scaling at the critical point. This collapse is shown in Fig.~\ref{fig:r_par_collapse-Luitz}. Here, instead of fixing $\phi_c=0$, we leave it as a free parameter, while fixing $W_c \simeq 3.8$ as results from the linear extrapolation in Fig.~\ref{fig:phiA_vs_W}. This fit yields the same exponent $\nu \simeq 0.9$ already obtained in \cite{Luitz15}, which violates the Harris bound. Moreover, it gives a value of $\phi_c \simeq 0.13$, which is inconsistent with the lowest finite-size critical value for the turning point $\phi_A = 0.085(4)$ observed for $W=4$.

In the inset of Fig.~\ref{fig:r_par_collapse-Luitz} we show the crossing point $\phi_c$ between data for the $r$-parameter at adjacent sizes as a function of the disorder strength $W$. One can observe as the crossing point is drifting towards $\phi_c = 0$, instead of saturating to a finite value.  

\newpage
\bibliography{references}

\providecommand{\noopsort}[1]{}\providecommand{\singleletter}[1]{#1}%
\begin{thebibliography}{88}%
\makeatletter
\providecommand \@ifxundefined [1]{%
 \@ifx{#1\undefined}
}%
\providecommand \@ifnum [1]{%
 \ifnum #1\expandafter \@firstoftwo
 \else \expandafter \@secondoftwo
 \fi
}%
\providecommand \@ifx [1]{%
 \ifx #1\expandafter \@firstoftwo
 \else \expandafter \@secondoftwo
 \fi
}%
\providecommand \natexlab [1]{#1}%
\providecommand \enquote  [1]{``#1''}%
\providecommand \bibnamefont  [1]{#1}%
\providecommand \bibfnamefont [1]{#1}%
\providecommand \citenamefont [1]{#1}%
\providecommand \href@noop [0]{\@secondoftwo}%
\providecommand \href [0]{\begingroup \@sanitize@url \@href}%
\providecommand \@href[1]{\@@startlink{#1}\@@href}%
\providecommand \@@href[1]{\endgroup#1\@@endlink}%
\providecommand \@sanitize@url [0]{\catcode `\\12\catcode `\$12\catcode `\&12\catcode `\#12\catcode `\^12\catcode `\_12\catcode `\%12\relax}%
\providecommand \@@startlink[1]{}%
\providecommand \@@endlink[0]{}%
\providecommand \url  [0]{\begingroup\@sanitize@url \@url }%
\providecommand \@url [1]{\endgroup\@href {#1}{\urlprefix }}%
\providecommand \urlprefix  [0]{URL }%
\providecommand \Eprint [0]{\href }%
\providecommand \doibase [0]{https://doi.org/}%
\providecommand \selectlanguage [0]{\@gobble}%
\providecommand \bibinfo  [0]{\@secondoftwo}%
\providecommand \bibfield  [0]{\@secondoftwo}%
\providecommand \translation [1]{[#1]}%
\providecommand \BibitemOpen [0]{}%
\providecommand \bibitemStop [0]{}%
\providecommand \bibitemNoStop [0]{.\EOS\space}%
\providecommand \EOS [0]{\spacefactor3000\relax}%
\providecommand \BibitemShut  [1]{\csname bibitem#1\endcsname}%
\let\auto@bib@innerbib\@empty
\bibitem [{\citenamefont {Edwards}\ and\ \citenamefont {Anderson}(1975)}]{edwards1975theory}%
  \BibitemOpen
  \bibfield  {author} {\bibinfo {author} {\bibfnamefont {S.~F.}\ \bibnamefont {Edwards}}\ and\ \bibinfo {author} {\bibfnamefont {P.~W.}\ \bibnamefont {Anderson}},\ }\bibfield  {title} {\bibinfo {title} {Theory of spin glasses},\ }\href@noop {} {\bibfield  {journal} {\bibinfo  {journal} {Journal of Physics F: Metal Physics}\ }\textbf {\bibinfo {volume} {5}},\ \bibinfo {pages} {965} (\bibinfo {year} {1975})}\BibitemShut {NoStop}%
\bibitem [{\citenamefont {Binder}\ and\ \citenamefont {Young}(1986)}]{binder1986spin}%
  \BibitemOpen
  \bibfield  {author} {\bibinfo {author} {\bibfnamefont {K.}~\bibnamefont {Binder}}\ and\ \bibinfo {author} {\bibfnamefont {A.~P.}\ \bibnamefont {Young}},\ }\bibfield  {title} {\bibinfo {title} {Spin glasses: Experimental facts, theoretical concepts, and open questions},\ }\href {https://dx.doi.org/10.1088/0305-4608/5/5/017} {\bibfield  {journal} {\bibinfo  {journal} {Reviews of Modern physics}\ }\textbf {\bibinfo {volume} {58}},\ \bibinfo {pages} {801} (\bibinfo {year} {1986})}\BibitemShut {NoStop}%
\bibitem [{\citenamefont {Mezard}\ \emph {et~al.}(1987)\citenamefont {Mezard}, \citenamefont {Parisi},\ and\ \citenamefont {Virasoro}}]{mezard1987spin}%
  \BibitemOpen
  \bibfield  {author} {\bibinfo {author} {\bibfnamefont {M.}~\bibnamefont {Mezard}}, \bibinfo {author} {\bibfnamefont {G.}~\bibnamefont {Parisi}},\ and\ \bibinfo {author} {\bibfnamefont {M.}~\bibnamefont {Virasoro}},\ }\href {https://books.google.it/books?id=DwY8DQAAQBAJ} {\emph {\bibinfo {title} {Spin Glass Theory And Beyond: An Introduction To The Replica Method And Its Applications}}},\ World Scientific Lecture Notes In Physics\ (\bibinfo  {publisher} {World Scientific Publishing Company},\ \bibinfo {year} {1987})\BibitemShut {NoStop}%
\bibitem [{\citenamefont {Anderson}(1958)}]{Anderson1958absence}%
  \BibitemOpen
  \bibfield  {author} {\bibinfo {author} {\bibfnamefont {P.~W.}\ \bibnamefont {Anderson}},\ }\bibfield  {title} {\bibinfo {title} {{Absence of Diffusion in Certain Random Lattices}},\ }\href {https://doi.org/10.1103/PhysRev.109.1492} {\bibfield  {journal} {\bibinfo  {journal} {Phys. Rev.}\ }\textbf {\bibinfo {volume} {109}},\ \bibinfo {pages} {1492} (\bibinfo {year} {1958})}\BibitemShut {NoStop}%
\bibitem [{\citenamefont {Evers}\ and\ \citenamefont {Mirlin}(2008)}]{evers2008anderson}%
  \BibitemOpen
  \bibfield  {author} {\bibinfo {author} {\bibfnamefont {F.}~\bibnamefont {Evers}}\ and\ \bibinfo {author} {\bibfnamefont {A.~D.}\ \bibnamefont {Mirlin}},\ }\bibfield  {title} {\bibinfo {title} {Anderson transitions},\ }\href@noop {} {\bibfield  {journal} {\bibinfo  {journal} {Reviews of Modern Physics}\ }\textbf {\bibinfo {volume} {80}},\ \bibinfo {pages} {1355} (\bibinfo {year} {2008})}\BibitemShut {NoStop}%
\bibitem [{\citenamefont {Roati}\ \emph {et~al.}(2008)\citenamefont {Roati}, \citenamefont {D’Errico}, \citenamefont {Fallani}, \citenamefont {Fattori}, \citenamefont {Fort}, \citenamefont {Zaccanti}, \citenamefont {Modugno}, \citenamefont {Modugno},\ and\ \citenamefont {Inguscio}}]{roati2008anderson}%
  \BibitemOpen
  \bibfield  {author} {\bibinfo {author} {\bibfnamefont {G.}~\bibnamefont {Roati}}, \bibinfo {author} {\bibfnamefont {C.}~\bibnamefont {D’Errico}}, \bibinfo {author} {\bibfnamefont {L.}~\bibnamefont {Fallani}}, \bibinfo {author} {\bibfnamefont {M.}~\bibnamefont {Fattori}}, \bibinfo {author} {\bibfnamefont {C.}~\bibnamefont {Fort}}, \bibinfo {author} {\bibfnamefont {M.}~\bibnamefont {Zaccanti}}, \bibinfo {author} {\bibfnamefont {G.}~\bibnamefont {Modugno}}, \bibinfo {author} {\bibfnamefont {M.}~\bibnamefont {Modugno}},\ and\ \bibinfo {author} {\bibfnamefont {M.}~\bibnamefont {Inguscio}},\ }\bibfield  {title} {\bibinfo {title} {{A}nderson localization of a non-interacting {B}ose--{E}instein condensate},\ }\href@noop {} {\bibfield  {journal} {\bibinfo  {journal} {Nature}\ }\textbf {\bibinfo {volume} {453}},\ \bibinfo {pages} {895} (\bibinfo {year} {2008})}\BibitemShut {NoStop}%
\bibitem [{\citenamefont {Billy}\ \emph {et~al.}(2008)\citenamefont {Billy}, \citenamefont {Josse}, \citenamefont {Zuo}, \citenamefont {Bernard}, \citenamefont {Hambrecht}, \citenamefont {Lugan}, \citenamefont {Cl{\'e}ment}, \citenamefont {Sanchez-Palencia}, \citenamefont {Bouyer},\ and\ \citenamefont {Aspect}}]{billy2008direct}%
  \BibitemOpen
  \bibfield  {author} {\bibinfo {author} {\bibfnamefont {J.}~\bibnamefont {Billy}}, \bibinfo {author} {\bibfnamefont {V.}~\bibnamefont {Josse}}, \bibinfo {author} {\bibfnamefont {Z.}~\bibnamefont {Zuo}}, \bibinfo {author} {\bibfnamefont {A.}~\bibnamefont {Bernard}}, \bibinfo {author} {\bibfnamefont {B.}~\bibnamefont {Hambrecht}}, \bibinfo {author} {\bibfnamefont {P.}~\bibnamefont {Lugan}}, \bibinfo {author} {\bibfnamefont {D.}~\bibnamefont {Cl{\'e}ment}}, \bibinfo {author} {\bibfnamefont {L.}~\bibnamefont {Sanchez-Palencia}}, \bibinfo {author} {\bibfnamefont {P.}~\bibnamefont {Bouyer}},\ and\ \bibinfo {author} {\bibfnamefont {A.}~\bibnamefont {Aspect}},\ }\bibfield  {title} {\bibinfo {title} {Direct observation of {A}nderson localization of matter waves in a controlled disorder},\ }\href@noop {} {\bibfield  {journal} {\bibinfo  {journal} {Nature}\ }\textbf {\bibinfo {volume} {453}},\ \bibinfo {pages} {891} (\bibinfo {year} {2008})}\BibitemShut {NoStop}%
\bibitem [{\citenamefont {Fleishman}\ and\ \citenamefont {Anderson}(1980)}]{fleishman1980interactions}%
  \BibitemOpen
  \bibfield  {author} {\bibinfo {author} {\bibfnamefont {L.}~\bibnamefont {Fleishman}}\ and\ \bibinfo {author} {\bibfnamefont {P.~W.}\ \bibnamefont {Anderson}},\ }\bibfield  {title} {\bibinfo {title} {Interactions and the {A}nderson transition},\ }\href@noop {} {\bibfield  {journal} {\bibinfo  {journal} {Physical Review B}\ }\textbf {\bibinfo {volume} {21}},\ \bibinfo {pages} {2366} (\bibinfo {year} {1980})}\BibitemShut {NoStop}%
\bibitem [{\citenamefont {Finkelstein}(1983)}]{finkelstein_83}%
  \BibitemOpen
  \bibfield  {author} {\bibinfo {author} {\bibfnamefont {A.~M.}\ \bibnamefont {Finkelstein}},\ }\bibfield  {title} {\bibinfo {title} {Influence of {C}oulomb interaction on the properties of disordered metals},\ }\href {http://www.jetp.ac.ru/cgi-bin/e/index/e/57/1/p97?a=list} {\bibfield  {journal} {\bibinfo  {journal} {Zh. Eksp. Teor. Fiz.}\ }\textbf {\bibinfo {volume} {168}} (\bibinfo {year} {1983})}\BibitemShut {NoStop}%
\bibitem [{\citenamefont {Castellani}\ \emph {et~al.}(1984{\natexlab{a}})\citenamefont {Castellani}, \citenamefont {Di~Castro}, \citenamefont {Lee},\ and\ \citenamefont {Ma}}]{castellani_1984}%
  \BibitemOpen
  \bibfield  {author} {\bibinfo {author} {\bibfnamefont {C.}~\bibnamefont {Castellani}}, \bibinfo {author} {\bibfnamefont {C.}~\bibnamefont {Di~Castro}}, \bibinfo {author} {\bibfnamefont {P.~A.}\ \bibnamefont {Lee}},\ and\ \bibinfo {author} {\bibfnamefont {M.}~\bibnamefont {Ma}},\ }\bibfield  {title} {\bibinfo {title} {{Interaction-driven metal-insulator transitions in disordered fermion systems}},\ }\href {https://doi.org/10.1103/PhysRevB.30.527} {\bibfield  {journal} {\bibinfo  {journal} {Phys. Rev. B}\ }\textbf {\bibinfo {volume} {30}},\ \bibinfo {pages} {527} (\bibinfo {year} {1984}{\natexlab{a}})}\BibitemShut {NoStop}%
\bibitem [{\citenamefont {Castellani}\ \emph {et~al.}(1984{\natexlab{b}})\citenamefont {Castellani}, \citenamefont {di~Castro},\ and\ \citenamefont {Forgacs}}]{castellani_1984_2}%
  \BibitemOpen
  \bibfield  {author} {\bibinfo {author} {\bibfnamefont {C.}~\bibnamefont {Castellani}}, \bibinfo {author} {\bibfnamefont {C.}~\bibnamefont {di~Castro}},\ and\ \bibinfo {author} {\bibfnamefont {G.}~\bibnamefont {Forgacs}},\ }\bibfield  {title} {\bibinfo {title} {{Renormalizability of the density of states of interacting disordered electron system}},\ }\href {https://doi.org/10.1103/PhysRevB.30.1593} {\bibfield  {journal} {\bibinfo  {journal} {Phys. Rev. B}\ }\textbf {\bibinfo {volume} {30}},\ \bibinfo {pages} {1593} (\bibinfo {year} {1984}{\natexlab{b}})}\BibitemShut {NoStop}%
\bibitem [{\citenamefont {Sierant}\ \emph {et~al.}(2025)\citenamefont {Sierant}, \citenamefont {Lewenstein}, \citenamefont {Scardicchio}, \citenamefont {Vidmar},\ and\ \citenamefont {Zakrzewski}}]{sierant24MBLreview}%
  \BibitemOpen
  \bibfield  {author} {\bibinfo {author} {\bibfnamefont {P.}~\bibnamefont {Sierant}}, \bibinfo {author} {\bibfnamefont {M.}~\bibnamefont {Lewenstein}}, \bibinfo {author} {\bibfnamefont {A.}~\bibnamefont {Scardicchio}}, \bibinfo {author} {\bibfnamefont {L.}~\bibnamefont {Vidmar}},\ and\ \bibinfo {author} {\bibfnamefont {J.}~\bibnamefont {Zakrzewski}},\ }\bibfield  {title} {\bibinfo {title} {Many-body localization in the age of classical computing*},\ }\href {https://doi.org/10.1088/1361-6633/ad9756} {\bibfield  {journal} {\bibinfo  {journal} {Rep. Progr. Phys.}\ }\textbf {\bibinfo {volume} {88}},\ \bibinfo {pages} {026502} (\bibinfo {year} {2025})}\BibitemShut {NoStop}%
\bibitem [{\citenamefont {Gornyi}\ \emph {et~al.}(2005)\citenamefont {Gornyi}, \citenamefont {Mirlin},\ and\ \citenamefont {Polyakov}}]{Gornyi2005Interacting}%
  \BibitemOpen
  \bibfield  {author} {\bibinfo {author} {\bibfnamefont {I.~V.}\ \bibnamefont {Gornyi}}, \bibinfo {author} {\bibfnamefont {A.~D.}\ \bibnamefont {Mirlin}},\ and\ \bibinfo {author} {\bibfnamefont {D.~G.}\ \bibnamefont {Polyakov}},\ }\bibfield  {title} {\bibinfo {title} {Interacting electrons in disordered wires: Anderson localization and low-$t$ transport},\ }\href {https://doi.org/10.1103/PhysRevLett.95.206603} {\bibfield  {journal} {\bibinfo  {journal} {Phys. Rev. Lett.}\ }\textbf {\bibinfo {volume} {95}},\ \bibinfo {pages} {206603} (\bibinfo {year} {2005})}\BibitemShut {NoStop}%
\bibitem [{\citenamefont {Basko}\ \emph {et~al.}(2006)\citenamefont {Basko}, \citenamefont {Aleiner},\ and\ \citenamefont {Altshuler}}]{Basko06}%
  \BibitemOpen
  \bibfield  {author} {\bibinfo {author} {\bibfnamefont {D.}~\bibnamefont {Basko}}, \bibinfo {author} {\bibfnamefont {I.}~\bibnamefont {Aleiner}},\ and\ \bibinfo {author} {\bibfnamefont {B.}~\bibnamefont {Altshuler}},\ }\bibfield  {title} {\bibinfo {title} {Metal{\textendash}insulator transition in a weakly interacting many-electron system with localized single-particle states},\ }\href {https://doi.org/10.1016/j.aop.2005.11.014} {\bibfield  {journal} {\bibinfo  {journal} {Ann. Phys.}\ }\textbf {\bibinfo {volume} {321}},\ \bibinfo {pages} {1126} (\bibinfo {year} {2006})}\BibitemShut {NoStop}%
\bibitem [{\citenamefont {Altshuler}\ \emph {et~al.}(1997)\citenamefont {Altshuler}, \citenamefont {Gefen}, \citenamefont {Kamenev},\ and\ \citenamefont {Levitov}}]{altshuler1997quasiparticle}%
  \BibitemOpen
  \bibfield  {author} {\bibinfo {author} {\bibfnamefont {B.~L.}\ \bibnamefont {Altshuler}}, \bibinfo {author} {\bibfnamefont {Y.}~\bibnamefont {Gefen}}, \bibinfo {author} {\bibfnamefont {A.}~\bibnamefont {Kamenev}},\ and\ \bibinfo {author} {\bibfnamefont {L.~S.}\ \bibnamefont {Levitov}},\ }\bibfield  {title} {\bibinfo {title} {Quasiparticle lifetime in a finite system: A nonperturbative approach},\ }\href {https://doi.org/10.1103/PhysRevLett.78.2803} {\bibfield  {journal} {\bibinfo  {journal} {Phys. Rev. Lett.}\ }\textbf {\bibinfo {volume} {78}},\ \bibinfo {pages} {2803} (\bibinfo {year} {1997})}\BibitemShut {NoStop}%
\bibitem [{\citenamefont {Oganesyan}\ and\ \citenamefont {Huse}(2007)}]{oganesyan2007localization}%
  \BibitemOpen
  \bibfield  {author} {\bibinfo {author} {\bibfnamefont {V.}~\bibnamefont {Oganesyan}}\ and\ \bibinfo {author} {\bibfnamefont {D.~A.}\ \bibnamefont {Huse}},\ }\bibfield  {title} {\bibinfo {title} {Localization of interacting fermions at high temperature},\ }\href@noop {} {\bibfield  {journal} {\bibinfo  {journal} {Physical review b}\ }\textbf {\bibinfo {volume} {75}},\ \bibinfo {pages} {155111} (\bibinfo {year} {2007})}\BibitemShut {NoStop}%
\bibitem [{\citenamefont {\v{Z}nidari\v{c}}\ \emph {et~al.}(2008)\citenamefont {\v{Z}nidari\v{c}}, \citenamefont {Prosen},\ and\ \citenamefont {Prelov\v{s}ek}}]{Znidaric2008Many}%
  \BibitemOpen
  \bibfield  {author} {\bibinfo {author} {\bibfnamefont {M.}~\bibnamefont {\v{Z}nidari\v{c}}}, \bibinfo {author} {\bibfnamefont {T.}~\bibnamefont {Prosen}},\ and\ \bibinfo {author} {\bibfnamefont {P.}~\bibnamefont {Prelov\v{s}ek}},\ }\bibfield  {title} {\bibinfo {title} {{Many-body localization in the Heisenberg XXZ magnet in a random field}},\ }\href {https://doi.org/10.1103/PhysRevB.77.064426} {\bibfield  {journal} {\bibinfo  {journal} {Phys. Rev. B}\ }\textbf {\bibinfo {volume} {77}},\ \bibinfo {pages} {064426} (\bibinfo {year} {2008})}\BibitemShut {NoStop}%
\bibitem [{\citenamefont {Pal}\ and\ \citenamefont {Huse}(2010)}]{Pal10}%
  \BibitemOpen
  \bibfield  {author} {\bibinfo {author} {\bibfnamefont {A.}~\bibnamefont {Pal}}\ and\ \bibinfo {author} {\bibfnamefont {D.~A.}\ \bibnamefont {Huse}},\ }\bibfield  {title} {\bibinfo {title} {Many-body localization phase transition},\ }\href {https://doi.org/10.1103/PhysRevB.82.174411} {\bibfield  {journal} {\bibinfo  {journal} {Phys. Rev. B}\ }\textbf {\bibinfo {volume} {82}},\ \bibinfo {pages} {174411} (\bibinfo {year} {2010})}\BibitemShut {NoStop}%
\bibitem [{\citenamefont {Bardarson}\ \emph {et~al.}(2012)\citenamefont {Bardarson}, \citenamefont {Pollmann},\ and\ \citenamefont {Moore}}]{Bardarson2012Unbounded}%
  \BibitemOpen
  \bibfield  {author} {\bibinfo {author} {\bibfnamefont {J.~H.}\ \bibnamefont {Bardarson}}, \bibinfo {author} {\bibfnamefont {F.}~\bibnamefont {Pollmann}},\ and\ \bibinfo {author} {\bibfnamefont {J.~E.}\ \bibnamefont {Moore}},\ }\bibfield  {title} {\bibinfo {title} {Unbounded growth of entanglement in models of many-body localization},\ }\href {https://doi.org/10.1103/PhysRevLett.109.017202} {\bibfield  {journal} {\bibinfo  {journal} {Phys. Rev. Lett.}\ }\textbf {\bibinfo {volume} {109}},\ \bibinfo {pages} {017202} (\bibinfo {year} {2012})}\BibitemShut {NoStop}%
\bibitem [{\citenamefont {Luca}\ and\ \citenamefont {Scardicchio}(2013)}]{Deluca13}%
  \BibitemOpen
  \bibfield  {author} {\bibinfo {author} {\bibfnamefont {A.~D.}\ \bibnamefont {Luca}}\ and\ \bibinfo {author} {\bibfnamefont {A.}~\bibnamefont {Scardicchio}},\ }\bibfield  {title} {\bibinfo {title} {Ergodicity breaking in a model showing many-body localization},\ }\href {https://doi.org/10.1209/0295-5075/101/37003} {\bibfield  {journal} {\bibinfo  {journal} {Europhys. Lett.}\ }\textbf {\bibinfo {volume} {101}},\ \bibinfo {pages} {37003} (\bibinfo {year} {2013})}\BibitemShut {NoStop}%
\bibitem [{\citenamefont {Luitz}\ \emph {et~al.}(2015)\citenamefont {Luitz}, \citenamefont {Laflorencie},\ and\ \citenamefont {Alet}}]{Luitz15}%
  \BibitemOpen
  \bibfield  {author} {\bibinfo {author} {\bibfnamefont {D.~J.}\ \bibnamefont {Luitz}}, \bibinfo {author} {\bibfnamefont {N.}~\bibnamefont {Laflorencie}},\ and\ \bibinfo {author} {\bibfnamefont {F.}~\bibnamefont {Alet}},\ }\bibfield  {title} {\bibinfo {title} {Many-body localization edge in the random-field {H}eisenberg chain},\ }\href {https://doi.org/10.1103/PhysRevB.91.081103} {\bibfield  {journal} {\bibinfo  {journal} {Phys. Rev. B}\ }\textbf {\bibinfo {volume} {91}},\ \bibinfo {pages} {081103} (\bibinfo {year} {2015})}\BibitemShut {NoStop}%
\bibitem [{\citenamefont {Pietracaprina}\ \emph {et~al.}(2018)\citenamefont {Pietracaprina}, \citenamefont {Macé}, \citenamefont {Luitz},\ and\ \citenamefont {Alet}}]{Pietracaprina2018Shift}%
  \BibitemOpen
  \bibfield  {author} {\bibinfo {author} {\bibfnamefont {F.}~\bibnamefont {Pietracaprina}}, \bibinfo {author} {\bibfnamefont {N.}~\bibnamefont {Macé}}, \bibinfo {author} {\bibfnamefont {D.~J.}\ \bibnamefont {Luitz}},\ and\ \bibinfo {author} {\bibfnamefont {F.}~\bibnamefont {Alet}},\ }\bibfield  {title} {\bibinfo {title} {{Shift-invert diagonalization of large many-body localizing spin chains}},\ }\href {https://doi.org/10.21468/SciPostPhys.5.5.045} {\bibfield  {journal} {\bibinfo  {journal} {SciPost Phys.}\ }\textbf {\bibinfo {volume} {5}},\ \bibinfo {pages} {045} (\bibinfo {year} {2018})}\BibitemShut {NoStop}%
\bibitem [{\citenamefont {Sierant}\ \emph {et~al.}(2020)\citenamefont {Sierant}, \citenamefont {Lewenstein},\ and\ \citenamefont {Zakrzewski}}]{Sierant2020Polynomially}%
  \BibitemOpen
  \bibfield  {author} {\bibinfo {author} {\bibfnamefont {P.}~\bibnamefont {Sierant}}, \bibinfo {author} {\bibfnamefont {M.}~\bibnamefont {Lewenstein}},\ and\ \bibinfo {author} {\bibfnamefont {J.}~\bibnamefont {Zakrzewski}},\ }\bibfield  {title} {\bibinfo {title} {Polynomially filtered exact diagonalization approach to many-body localization},\ }\href {https://doi.org/10.1103/PhysRevLett.125.156601} {\bibfield  {journal} {\bibinfo  {journal} {Phys. Rev. Lett.}\ }\textbf {\bibinfo {volume} {125}},\ \bibinfo {pages} {156601} (\bibinfo {year} {2020})}\BibitemShut {NoStop}%
\bibitem [{\citenamefont {Colbois}\ \emph {et~al.}(2024{\natexlab{a}})\citenamefont {Colbois}, \citenamefont {Alet},\ and\ \citenamefont {Laflorencie}}]{Colbois2024prl}%
  \BibitemOpen
  \bibfield  {author} {\bibinfo {author} {\bibfnamefont {J.}~\bibnamefont {Colbois}}, \bibinfo {author} {\bibfnamefont {F.}~\bibnamefont {Alet}},\ and\ \bibinfo {author} {\bibfnamefont {N.}~\bibnamefont {Laflorencie}},\ }\bibfield  {title} {\bibinfo {title} {Interaction-driven instabilities in the random-field {XXZ} chain},\ }\href {https://doi.org/10.1103/PhysRevLett.133.116502} {\bibfield  {journal} {\bibinfo  {journal} {Phys. Rev. Lett.}\ }\textbf {\bibinfo {volume} {133}},\ \bibinfo {pages} {116502} (\bibinfo {year} {2024}{\natexlab{a}})}\BibitemShut {NoStop}%
\bibitem [{\citenamefont {Colbois}\ \emph {et~al.}(2024{\natexlab{b}})\citenamefont {Colbois}, \citenamefont {Alet},\ and\ \citenamefont {Laflorencie}}]{colbois2024b}%
  \BibitemOpen
  \bibfield  {author} {\bibinfo {author} {\bibfnamefont {J.}~\bibnamefont {Colbois}}, \bibinfo {author} {\bibfnamefont {F.}~\bibnamefont {Alet}},\ and\ \bibinfo {author} {\bibfnamefont {N.}~\bibnamefont {Laflorencie}},\ }\bibfield  {title} {\bibinfo {title} {{Statistics of systemwide correlations in the random-field XXZ chain: Importance of rare events in the many-body localized phase}},\ }\bibfield  {journal} {\bibinfo  {journal} {Physical Review B}\ }\textbf {\bibinfo {volume} {110}},\ \href {https://doi.org/10.1103/physrevb.110.214210} {10.1103/physrevb.110.214210} (\bibinfo {year} {2024}{\natexlab{b}})\BibitemShut {NoStop}%
\bibitem [{\citenamefont {Serbyn}\ \emph {et~al.}(2013)\citenamefont {Serbyn}, \citenamefont {Papi{\'c}},\ and\ \citenamefont {Abanin}}]{serbyn2013local}%
  \BibitemOpen
  \bibfield  {author} {\bibinfo {author} {\bibfnamefont {M.}~\bibnamefont {Serbyn}}, \bibinfo {author} {\bibfnamefont {Z.}~\bibnamefont {Papi{\'c}}},\ and\ \bibinfo {author} {\bibfnamefont {D.~A.}\ \bibnamefont {Abanin}},\ }\bibfield  {title} {\bibinfo {title} {Local conservation laws and the structure of the many-body localized states},\ }\href@noop {} {\bibfield  {journal} {\bibinfo  {journal} {Physical review letters}\ }\textbf {\bibinfo {volume} {111}},\ \bibinfo {pages} {127201} (\bibinfo {year} {2013})}\BibitemShut {NoStop}%
\bibitem [{\citenamefont {Huse}\ \emph {et~al.}(2014)\citenamefont {Huse}, \citenamefont {Nandkishore},\ and\ \citenamefont {Oganesyan}}]{Huse2014Phenomenology}%
  \BibitemOpen
  \bibfield  {author} {\bibinfo {author} {\bibfnamefont {D.~A.}\ \bibnamefont {Huse}}, \bibinfo {author} {\bibfnamefont {R.}~\bibnamefont {Nandkishore}},\ and\ \bibinfo {author} {\bibfnamefont {V.}~\bibnamefont {Oganesyan}},\ }\bibfield  {title} {\bibinfo {title} {{Phenomenology of fully many-body-localized systems}},\ }\href {https://doi.org/10.1103/PhysRevB.90.174202} {\bibfield  {journal} {\bibinfo  {journal} {Phys. Rev. B}\ }\textbf {\bibinfo {volume} {90}},\ \bibinfo {pages} {174202} (\bibinfo {year} {2014})}\BibitemShut {NoStop}%
\bibitem [{\citenamefont {Ros}\ \emph {et~al.}(2015)\citenamefont {Ros}, \citenamefont {Müller},\ and\ \citenamefont {Scardicchio}}]{ros2015integrals}%
  \BibitemOpen
  \bibfield  {author} {\bibinfo {author} {\bibfnamefont {V.}~\bibnamefont {Ros}}, \bibinfo {author} {\bibfnamefont {M.}~\bibnamefont {Müller}},\ and\ \bibinfo {author} {\bibfnamefont {A.}~\bibnamefont {Scardicchio}},\ }\bibfield  {title} {\bibinfo {title} {Integrals of motion in the many-body localized phase},\ }\href {https://doi.org/10.1016/j.nuclphysb.2014.12.014} {\bibfield  {journal} {\bibinfo  {journal} {Nucl. Phys. B}\ }\textbf {\bibinfo {volume} {891}},\ \bibinfo {pages} {420} (\bibinfo {year} {2015})}\BibitemShut {NoStop}%
\bibitem [{\citenamefont {Imbrie}(2016{\natexlab{a}})}]{Imbrie2016Many}%
  \BibitemOpen
  \bibfield  {author} {\bibinfo {author} {\bibfnamefont {J.~Z.}\ \bibnamefont {Imbrie}},\ }\bibfield  {title} {\bibinfo {title} {On many-body localization for quantum spin chains},\ }\href {https://doi.org/10.1007/s10955-016-1508-x} {\bibfield  {journal} {\bibinfo  {journal} {J. Stat. Phys.}\ }\textbf {\bibinfo {volume} {163}},\ \bibinfo {pages} {998} (\bibinfo {year} {2016}{\natexlab{a}})}\BibitemShut {NoStop}%
\bibitem [{\citenamefont {Imbrie}(2016{\natexlab{b}})}]{Imbrie2016Diagonalization}%
  \BibitemOpen
  \bibfield  {author} {\bibinfo {author} {\bibfnamefont {J.~Z.}\ \bibnamefont {Imbrie}},\ }\bibfield  {title} {\bibinfo {title} {Diagonalization and many-body localization for a disordered quantum spin chain},\ }\href {https://doi.org/10.1103/PhysRevLett.117.027201} {\bibfield  {journal} {\bibinfo  {journal} {Phys. Rev. Lett.}\ }\textbf {\bibinfo {volume} {117}},\ \bibinfo {pages} {027201} (\bibinfo {year} {2016}{\natexlab{b}})}\BibitemShut {NoStop}%
\bibitem [{\citenamefont {Imbrie}\ \emph {et~al.}(2017{\natexlab{a}})\citenamefont {Imbrie}, \citenamefont {Ros},\ and\ \citenamefont {Scardicchio}}]{Imbrie17}%
  \BibitemOpen
  \bibfield  {author} {\bibinfo {author} {\bibfnamefont {J.~Z.}\ \bibnamefont {Imbrie}}, \bibinfo {author} {\bibfnamefont {V.}~\bibnamefont {Ros}},\ and\ \bibinfo {author} {\bibfnamefont {A.}~\bibnamefont {Scardicchio}},\ }\bibfield  {title} {\bibinfo {title} {Local integrals of motion in many-body localized systems},\ }\href {https://doi.org/https://doi.org/10.1002/andp.201600278} {\bibfield  {journal} {\bibinfo  {journal} {Ann. Phys.}\ }\textbf {\bibinfo {volume} {529}},\ \bibinfo {pages} {1600278} (\bibinfo {year} {2017}{\natexlab{a}})}\BibitemShut {NoStop}%
\bibitem [{\citenamefont {De~Roeck}\ \emph {et~al.}(2024)\citenamefont {De~Roeck}, \citenamefont {Giacomin}, \citenamefont {Huveneers},\ and\ \citenamefont {Prosniak}}]{deroeck2024absence}%
  \BibitemOpen
  \bibfield  {author} {\bibinfo {author} {\bibfnamefont {W.}~\bibnamefont {De~Roeck}}, \bibinfo {author} {\bibfnamefont {L.}~\bibnamefont {Giacomin}}, \bibinfo {author} {\bibfnamefont {F.}~\bibnamefont {Huveneers}},\ and\ \bibinfo {author} {\bibfnamefont {O.}~\bibnamefont {Prosniak}},\ }\bibfield  {title} {\bibinfo {title} {Absence of normal heat conduction in strongly disordered interacting quantum chains},\ }\href@noop {} {\bibfield  {journal} {\bibinfo  {journal} {arXiv preprint arXiv:2408.04338}\ } (\bibinfo {year} {2024})}\BibitemShut {NoStop}%
\bibitem [{\citenamefont {Altshuler}\ \emph {et~al.}(2024)\citenamefont {Altshuler}, \citenamefont {Kravtsov}, \citenamefont {Scardicchio}, \citenamefont {Sierant},\ and\ \citenamefont {Vanoni}}]{altshuler2024renormalization}%
  \BibitemOpen
  \bibfield  {author} {\bibinfo {author} {\bibfnamefont {B.~L.}\ \bibnamefont {Altshuler}}, \bibinfo {author} {\bibfnamefont {V.~E.}\ \bibnamefont {Kravtsov}}, \bibinfo {author} {\bibfnamefont {A.}~\bibnamefont {Scardicchio}}, \bibinfo {author} {\bibfnamefont {P.}~\bibnamefont {Sierant}},\ and\ \bibinfo {author} {\bibfnamefont {C.}~\bibnamefont {Vanoni}},\ }\bibfield  {title} {\bibinfo {title} {{Renormalization group for Anderson localization on high-dimensional lattices}},\ }\href@noop {} {\bibfield  {journal} {\bibinfo  {journal} {arXiv}\ } (\bibinfo {year} {2024})},\ \Eprint {https://arxiv.org/abs/2403.01974} {arXiv:2403.01974} \BibitemShut {NoStop}%
\bibitem [{\citenamefont {Vanoni}\ \emph {et~al.}(2024)\citenamefont {Vanoni}, \citenamefont {Altshuler}, \citenamefont {Kravtsov},\ and\ \citenamefont {Scardicchio}}]{vanoni2023renormalization}%
  \BibitemOpen
  \bibfield  {author} {\bibinfo {author} {\bibfnamefont {C.}~\bibnamefont {Vanoni}}, \bibinfo {author} {\bibfnamefont {B.~L.}\ \bibnamefont {Altshuler}}, \bibinfo {author} {\bibfnamefont {V.~E.}\ \bibnamefont {Kravtsov}},\ and\ \bibinfo {author} {\bibfnamefont {A.}~\bibnamefont {Scardicchio}},\ }\bibfield  {title} {\bibinfo {title} {Renormalization group analysis of the {A}nderson model on random regular graphs},\ }\bibfield  {journal} {\bibinfo  {journal} {Proceedings of the National Academy of Sciences}\ }\textbf {\bibinfo {volume} {121}},\ \href {https://doi.org/10.1073/pnas.2401955121} {10.1073/pnas.2401955121} (\bibinfo {year} {2024})\BibitemShut {NoStop}%
\bibitem [{\citenamefont {Kutlin}\ and\ \citenamefont {Vanoni}(2024)}]{kutlin2024investigating}%
  \BibitemOpen
  \bibfield  {author} {\bibinfo {author} {\bibfnamefont {A.}~\bibnamefont {Kutlin}}\ and\ \bibinfo {author} {\bibfnamefont {C.}~\bibnamefont {Vanoni}},\ }\bibfield  {title} {\bibinfo {title} {{Investigating finite-size effects in random matrices by counting resonances}},\ }\href@noop {} {\bibfield  {journal} {\bibinfo  {journal} {arXiv}\ } (\bibinfo {year} {2024})},\ \Eprint {https://arxiv.org/abs/2402.10271} {arXiv:2402.10271} \BibitemShut {NoStop}%
\bibitem [{\citenamefont {Goremykina}\ \emph {et~al.}(2019)\citenamefont {Goremykina}, \citenamefont {Vasseur},\ and\ \citenamefont {Serbyn}}]{Goremykina2019Analytically}%
  \BibitemOpen
  \bibfield  {author} {\bibinfo {author} {\bibfnamefont {A.}~\bibnamefont {Goremykina}}, \bibinfo {author} {\bibfnamefont {R.}~\bibnamefont {Vasseur}},\ and\ \bibinfo {author} {\bibfnamefont {M.}~\bibnamefont {Serbyn}},\ }\bibfield  {title} {\bibinfo {title} {Analytically solvable renormalization group for the many-body localization transition},\ }\href {https://doi.org/10.1103/PhysRevLett.122.040601} {\bibfield  {journal} {\bibinfo  {journal} {Phys. Rev. Lett.}\ }\textbf {\bibinfo {volume} {122}},\ \bibinfo {pages} {040601} (\bibinfo {year} {2019})}\BibitemShut {NoStop}%
\bibitem [{\citenamefont {Dumitrescu}\ \emph {et~al.}(2019)\citenamefont {Dumitrescu}, \citenamefont {Goremykina}, \citenamefont {Parameswaran}, \citenamefont {Serbyn},\ and\ \citenamefont {Vasseur}}]{Dumitrescu2019Kosterlitz}%
  \BibitemOpen
  \bibfield  {author} {\bibinfo {author} {\bibfnamefont {P.~T.}\ \bibnamefont {Dumitrescu}}, \bibinfo {author} {\bibfnamefont {A.}~\bibnamefont {Goremykina}}, \bibinfo {author} {\bibfnamefont {S.~A.}\ \bibnamefont {Parameswaran}}, \bibinfo {author} {\bibfnamefont {M.}~\bibnamefont {Serbyn}},\ and\ \bibinfo {author} {\bibfnamefont {R.}~\bibnamefont {Vasseur}},\ }\bibfield  {title} {\bibinfo {title} {{Kosterlitz-Thouless scaling at many-body localization phase transitions}},\ }\href {https://doi.org/10.1103/PhysRevB.99.094205} {\bibfield  {journal} {\bibinfo  {journal} {Phys. Rev. B}\ }\textbf {\bibinfo {volume} {99}},\ \bibinfo {pages} {094205} (\bibinfo {year} {2019})}\BibitemShut {NoStop}%
\bibitem [{\citenamefont {Morningstar}\ and\ \citenamefont {Huse}(2019)}]{Morningstar2019Renormalization}%
  \BibitemOpen
  \bibfield  {author} {\bibinfo {author} {\bibfnamefont {A.}~\bibnamefont {Morningstar}}\ and\ \bibinfo {author} {\bibfnamefont {D.~A.}\ \bibnamefont {Huse}},\ }\bibfield  {title} {\bibinfo {title} {{Renormalization-group study of the many-body localization transition in one dimension}},\ }\href {https://doi.org/10.1103/PhysRevB.99.224205} {\bibfield  {journal} {\bibinfo  {journal} {Phys. Rev. B}\ }\textbf {\bibinfo {volume} {99}},\ \bibinfo {pages} {224205} (\bibinfo {year} {2019})}\BibitemShut {NoStop}%
\bibitem [{\citenamefont {Morningstar}\ \emph {et~al.}(2020)\citenamefont {Morningstar}, \citenamefont {Huse},\ and\ \citenamefont {Imbrie}}]{Morningstar2020Many}%
  \BibitemOpen
  \bibfield  {author} {\bibinfo {author} {\bibfnamefont {A.}~\bibnamefont {Morningstar}}, \bibinfo {author} {\bibfnamefont {D.~A.}\ \bibnamefont {Huse}},\ and\ \bibinfo {author} {\bibfnamefont {J.~Z.}\ \bibnamefont {Imbrie}},\ }\bibfield  {title} {\bibinfo {title} {Many-body localization near the critical point},\ }\href {https://doi.org/10.1103/PhysRevB.102.125134} {\bibfield  {journal} {\bibinfo  {journal} {Phys. Rev. B}\ }\textbf {\bibinfo {volume} {102}},\ \bibinfo {pages} {125134} (\bibinfo {year} {2020})}\BibitemShut {NoStop}%
\bibitem [{\citenamefont {\ifmmode \check{Z}\else \v{Z}\fi{}nidari\ifmmode~\check{c}\else \v{c}\fi{}}\ \emph {et~al.}(2016)\citenamefont {\ifmmode \check{Z}\else \v{Z}\fi{}nidari\ifmmode~\check{c}\else \v{c}\fi{}}, \citenamefont {Scardicchio},\ and\ \citenamefont {Varma}}]{vznidarivc2016diffusive}%
  \BibitemOpen
  \bibfield  {author} {\bibinfo {author} {\bibfnamefont {M.}~\bibnamefont {\ifmmode \check{Z}\else \v{Z}\fi{}nidari\ifmmode~\check{c}\else \v{c}\fi{}}}, \bibinfo {author} {\bibfnamefont {A.}~\bibnamefont {Scardicchio}},\ and\ \bibinfo {author} {\bibfnamefont {V.~K.}\ \bibnamefont {Varma}},\ }\bibfield  {title} {\bibinfo {title} {Diffusive and subdiffusive spin transport in the ergodic phase of a many-body localizable system},\ }\href {https://doi.org/10.1103/PhysRevLett.117.040601} {\bibfield  {journal} {\bibinfo  {journal} {Phys. Rev. Lett.}\ }\textbf {\bibinfo {volume} {117}},\ \bibinfo {pages} {040601} (\bibinfo {year} {2016})}\BibitemShut {NoStop}%
\bibitem [{\citenamefont {Doggen}\ \emph {et~al.}(2018)\citenamefont {Doggen}, \citenamefont {Schindler}, \citenamefont {Tikhonov}, \citenamefont {Mirlin}, \citenamefont {Neupert}, \citenamefont {Polyakov},\ and\ \citenamefont {Gornyi}}]{Diggen2018Manybody}%
  \BibitemOpen
  \bibfield  {author} {\bibinfo {author} {\bibfnamefont {E.~V.~H.}\ \bibnamefont {Doggen}}, \bibinfo {author} {\bibfnamefont {F.}~\bibnamefont {Schindler}}, \bibinfo {author} {\bibfnamefont {K.~S.}\ \bibnamefont {Tikhonov}}, \bibinfo {author} {\bibfnamefont {A.~D.}\ \bibnamefont {Mirlin}}, \bibinfo {author} {\bibfnamefont {T.}~\bibnamefont {Neupert}}, \bibinfo {author} {\bibfnamefont {D.~G.}\ \bibnamefont {Polyakov}},\ and\ \bibinfo {author} {\bibfnamefont {I.~V.}\ \bibnamefont {Gornyi}},\ }\bibfield  {title} {\bibinfo {title} {Many-body localization and delocalization in large quantum chains},\ }\href {https://doi.org/10.1103/PhysRevB.98.174202} {\bibfield  {journal} {\bibinfo  {journal} {Phys. Rev. B}\ }\textbf {\bibinfo {volume} {98}},\ \bibinfo {pages} {174202} (\bibinfo {year} {2018})}\BibitemShut {NoStop}%
\bibitem [{\citenamefont {Altman}\ and\ \citenamefont {Vosk}(2015)}]{Altman2015Universal}%
  \BibitemOpen
  \bibfield  {author} {\bibinfo {author} {\bibfnamefont {E.}~\bibnamefont {Altman}}\ and\ \bibinfo {author} {\bibfnamefont {R.}~\bibnamefont {Vosk}},\ }\bibfield  {title} {\bibinfo {title} {Universal dynamics and renormalization in many-body-localized systems},\ }\href {https://doi.org/10.1146/annurev-conmatphys-031214-014701} {\bibfield  {journal} {\bibinfo  {journal} {Annu. Rev. Condens. Matt. Phys.}\ }\textbf {\bibinfo {volume} {6}},\ \bibinfo {pages} {383} (\bibinfo {year} {2015})}\BibitemShut {NoStop}%
\bibitem [{\citenamefont {Nandkishore}\ and\ \citenamefont {Huse}(2015)}]{Nandkishore2015Many}%
  \BibitemOpen
  \bibfield  {author} {\bibinfo {author} {\bibfnamefont {R.}~\bibnamefont {Nandkishore}}\ and\ \bibinfo {author} {\bibfnamefont {D.~A.}\ \bibnamefont {Huse}},\ }\bibfield  {title} {\bibinfo {title} {Many-body localization and thermalization in quantum statistical mechanics},\ }\href {https://doi.org/10.1146/annurev-conmatphys-031214-014726} {\bibfield  {journal} {\bibinfo  {journal} {Annu. Rev. Condens. Matt. Phys.}\ }\textbf {\bibinfo {volume} {6}},\ \bibinfo {pages} {15} (\bibinfo {year} {2015})}\BibitemShut {NoStop}%
\bibitem [{\citenamefont {Alet}\ and\ \citenamefont {Laflorencie}(2018)}]{Alet2018Many}%
  \BibitemOpen
  \bibfield  {author} {\bibinfo {author} {\bibfnamefont {F.}~\bibnamefont {Alet}}\ and\ \bibinfo {author} {\bibfnamefont {N.}~\bibnamefont {Laflorencie}},\ }\bibfield  {title} {\bibinfo {title} {{Many-body localization: An introduction and selected topics}},\ }\href {https://doi.org/10.1016/j.crhy.2018.03.003} {\bibfield  {journal} {\bibinfo  {journal} {Comptes Rendus Physique}\ }\textbf {\bibinfo {volume} {19}},\ \bibinfo {pages} {498} (\bibinfo {year} {2018})}\BibitemShut {NoStop}%
\bibitem [{\citenamefont {Abanin}\ \emph {et~al.}(2019)\citenamefont {Abanin}, \citenamefont {Altman}, \citenamefont {Bloch},\ and\ \citenamefont {Serbyn}}]{Abanin2019colloquium}%
  \BibitemOpen
  \bibfield  {author} {\bibinfo {author} {\bibfnamefont {D.~A.}\ \bibnamefont {Abanin}}, \bibinfo {author} {\bibfnamefont {E.}~\bibnamefont {Altman}}, \bibinfo {author} {\bibfnamefont {I.}~\bibnamefont {Bloch}},\ and\ \bibinfo {author} {\bibfnamefont {M.}~\bibnamefont {Serbyn}},\ }\bibfield  {title} {\bibinfo {title} {{Colloquium: Many-body localization, thermalization, and entanglement}},\ }\href {https://doi.org/10.1103/RevModPhys.91.021001} {\bibfield  {journal} {\bibinfo  {journal} {Rev. Mod. Phys.}\ }\textbf {\bibinfo {volume} {91}},\ \bibinfo {pages} {021001} (\bibinfo {year} {2019})}\BibitemShut {NoStop}%
\bibitem [{\citenamefont {Gopalakrishnan}\ and\ \citenamefont {Parameswaran}(2020)}]{Gopalakrishnan2020Dynamics}%
  \BibitemOpen
  \bibfield  {author} {\bibinfo {author} {\bibfnamefont {S.}~\bibnamefont {Gopalakrishnan}}\ and\ \bibinfo {author} {\bibfnamefont {S.}~\bibnamefont {Parameswaran}},\ }\bibfield  {title} {\bibinfo {title} {{Dynamics and transport at the threshold of many-body localization}},\ }\href {https://doi.org/10.1016/j.physrep.2020.03.003} {\bibfield  {journal} {\bibinfo  {journal} {Phys. Rep.}\ }\textbf {\bibinfo {volume} {862}},\ \bibinfo {pages} {1} (\bibinfo {year} {2020})}\BibitemShut {NoStop}%
\bibitem [{\citenamefont {Imbrie}\ \emph {et~al.}(2017{\natexlab{b}})\citenamefont {Imbrie}, \citenamefont {Ros},\ and\ \citenamefont {Scardicchio}}]{Imbrie2016Local}%
  \BibitemOpen
  \bibfield  {author} {\bibinfo {author} {\bibfnamefont {J.~Z.}\ \bibnamefont {Imbrie}}, \bibinfo {author} {\bibfnamefont {V.}~\bibnamefont {Ros}},\ and\ \bibinfo {author} {\bibfnamefont {A.}~\bibnamefont {Scardicchio}},\ }\bibfield  {title} {\bibinfo {title} {{Local integrals of motion in many-body localized systems}},\ }\href {https://doi.org/10.1002/andp.201600278} {\bibfield  {journal} {\bibinfo  {journal} {Ann. Phys. (Berlin)}\ }\textbf {\bibinfo {volume} {529}},\ \bibinfo {pages} {1600278} (\bibinfo {year} {2017}{\natexlab{b}})}\BibitemShut {NoStop}%
\bibitem [{\citenamefont {De~Roeck}\ and\ \citenamefont {Huveneers}(2017)}]{DeRoeck2017Stability}%
  \BibitemOpen
  \bibfield  {author} {\bibinfo {author} {\bibfnamefont {W.}~\bibnamefont {De~Roeck}}\ and\ \bibinfo {author} {\bibfnamefont {F.}~\bibnamefont {Huveneers}},\ }\bibfield  {title} {\bibinfo {title} {{Stability and instability towards delocalization in many-body localization systems}},\ }\href {https://doi.org/10.1103/PhysRevB.95.155129} {\bibfield  {journal} {\bibinfo  {journal} {Phys. Rev. B}\ }\textbf {\bibinfo {volume} {95}},\ \bibinfo {pages} {155129} (\bibinfo {year} {2017})}\BibitemShut {NoStop}%
\bibitem [{\citenamefont {Thiery}\ \emph {et~al.}(2018)\citenamefont {Thiery}, \citenamefont {Huveneers}, \citenamefont {M{\"u}ller},\ and\ \citenamefont {De~Roeck}}]{Thiery2018Many}%
  \BibitemOpen
  \bibfield  {author} {\bibinfo {author} {\bibfnamefont {T.}~\bibnamefont {Thiery}}, \bibinfo {author} {\bibfnamefont {F.}~\bibnamefont {Huveneers}}, \bibinfo {author} {\bibfnamefont {M.}~\bibnamefont {M{\"u}ller}},\ and\ \bibinfo {author} {\bibfnamefont {W.}~\bibnamefont {De~Roeck}},\ }\bibfield  {title} {\bibinfo {title} {Many-body delocalization as a quantum avalanche},\ }\href {https://doi.org/10.1103/PhysRevLett.121.140601} {\bibfield  {journal} {\bibinfo  {journal} {Phys. Rev. Lett.}\ }\textbf {\bibinfo {volume} {121}},\ \bibinfo {pages} {140601} (\bibinfo {year} {2018})}\BibitemShut {NoStop}%
\bibitem [{\citenamefont {Ha}\ \emph {et~al.}(2023)\citenamefont {Ha}, \citenamefont {Morningstar},\ and\ \citenamefont {Huse}}]{Ha2023Manybody}%
  \BibitemOpen
  \bibfield  {author} {\bibinfo {author} {\bibfnamefont {H.}~\bibnamefont {Ha}}, \bibinfo {author} {\bibfnamefont {A.}~\bibnamefont {Morningstar}},\ and\ \bibinfo {author} {\bibfnamefont {D.~A.}\ \bibnamefont {Huse}},\ }\bibfield  {title} {\bibinfo {title} {Many-body resonances in the avalanche instability of many-body localization},\ }\href {https://doi.org/10.1103/PhysRevLett.130.250405} {\bibfield  {journal} {\bibinfo  {journal} {Phys. Rev. Lett.}\ }\textbf {\bibinfo {volume} {130}},\ \bibinfo {pages} {250405} (\bibinfo {year} {2023})}\BibitemShut {NoStop}%
\bibitem [{\citenamefont {Morningstar}\ \emph {et~al.}(2022)\citenamefont {Morningstar}, \citenamefont {Colmenarez}, \citenamefont {Khemani}, \citenamefont {Luitz},\ and\ \citenamefont {Huse}}]{Morningstar2022Avalanches}%
  \BibitemOpen
  \bibfield  {author} {\bibinfo {author} {\bibfnamefont {A.}~\bibnamefont {Morningstar}}, \bibinfo {author} {\bibfnamefont {L.}~\bibnamefont {Colmenarez}}, \bibinfo {author} {\bibfnamefont {V.}~\bibnamefont {Khemani}}, \bibinfo {author} {\bibfnamefont {D.~J.}\ \bibnamefont {Luitz}},\ and\ \bibinfo {author} {\bibfnamefont {D.~A.}\ \bibnamefont {Huse}},\ }\bibfield  {title} {\bibinfo {title} {Avalanches and many-body resonances in many-body localized systems},\ }\href {https://doi.org/10.1103/PhysRevB.105.174205} {\bibfield  {journal} {\bibinfo  {journal} {Phys. Rev. B}\ }\textbf {\bibinfo {volume} {105}},\ \bibinfo {pages} {174205} (\bibinfo {year} {2022})}\BibitemShut {NoStop}%
\bibitem [{\citenamefont {Chandran}\ \emph {et~al.}(2015)\citenamefont {Chandran}, \citenamefont {Laumann},\ and\ \citenamefont {Oganesyan}}]{Chandran2015Finite}%
  \BibitemOpen
  \bibfield  {author} {\bibinfo {author} {\bibfnamefont {A.}~\bibnamefont {Chandran}}, \bibinfo {author} {\bibfnamefont {C.~R.}\ \bibnamefont {Laumann}},\ and\ \bibinfo {author} {\bibfnamefont {V.}~\bibnamefont {Oganesyan}},\ }\href {https://arxiv.org/abs/1509.04285} {\bibinfo {title} {{Finite size scaling bounds on many-body localized phase transitions}}} (\bibinfo {year} {2015}),\ \Eprint {https://arxiv.org/abs/1509.04285} {arXiv:1509.04285} \BibitemShut {NoStop}%
\bibitem [{\citenamefont {Panda}\ \emph {et~al.}(2020)\citenamefont {Panda}, \citenamefont {Scardicchio}, \citenamefont {Schulz}, \citenamefont {Taylor},\ and\ \citenamefont {Žnidarič}}]{panda2020can}%
  \BibitemOpen
  \bibfield  {author} {\bibinfo {author} {\bibfnamefont {R.~K.}\ \bibnamefont {Panda}}, \bibinfo {author} {\bibfnamefont {A.}~\bibnamefont {Scardicchio}}, \bibinfo {author} {\bibfnamefont {M.}~\bibnamefont {Schulz}}, \bibinfo {author} {\bibfnamefont {S.~R.}\ \bibnamefont {Taylor}},\ and\ \bibinfo {author} {\bibfnamefont {M.}~\bibnamefont {Žnidarič}},\ }\bibfield  {title} {\bibinfo {title} {Can we study the many-body localisation transition?},\ }\href {https://doi.org/10.1209/0295-5075/128/67003} {\bibfield  {journal} {\bibinfo  {journal} {EPL (Europhysics Letters)}\ }\textbf {\bibinfo {volume} {128}},\ \bibinfo {pages} {67003} (\bibinfo {year} {2020})}\BibitemShut {NoStop}%
\bibitem [{\citenamefont {Abanin}\ \emph {et~al.}(2021)\citenamefont {Abanin}, \citenamefont {Bardarson}, \citenamefont {De~Tomasi}, \citenamefont {Gopalakrishnan}, \citenamefont {Khemani}, \citenamefont {Parameswaran}, \citenamefont {Pollmann}, \citenamefont {Potter}, \citenamefont {Serbyn},\ and\ \citenamefont {Vasseur}}]{abanin2021distinguishing}%
  \BibitemOpen
  \bibfield  {author} {\bibinfo {author} {\bibfnamefont {D.}~\bibnamefont {Abanin}}, \bibinfo {author} {\bibfnamefont {J.~H.}\ \bibnamefont {Bardarson}}, \bibinfo {author} {\bibfnamefont {G.}~\bibnamefont {De~Tomasi}}, \bibinfo {author} {\bibfnamefont {S.}~\bibnamefont {Gopalakrishnan}}, \bibinfo {author} {\bibfnamefont {V.}~\bibnamefont {Khemani}}, \bibinfo {author} {\bibfnamefont {S.}~\bibnamefont {Parameswaran}}, \bibinfo {author} {\bibfnamefont {F.}~\bibnamefont {Pollmann}}, \bibinfo {author} {\bibfnamefont {A.}~\bibnamefont {Potter}}, \bibinfo {author} {\bibfnamefont {M.}~\bibnamefont {Serbyn}},\ and\ \bibinfo {author} {\bibfnamefont {R.}~\bibnamefont {Vasseur}},\ }\bibfield  {title} {\bibinfo {title} {Distinguishing localization from chaos: Challenges in finite-size systems},\ }\href@noop {} {\bibfield  {journal} {\bibinfo  {journal} {Annals of Physics}\ }\textbf {\bibinfo {volume} {427}},\ \bibinfo {pages} {168415} (\bibinfo {year} {2021})}\BibitemShut {NoStop}%
\bibitem [{\citenamefont {Sierant}\ and\ \citenamefont {Zakrzewski}(2022)}]{sierant2022challenges}%
  \BibitemOpen
  \bibfield  {author} {\bibinfo {author} {\bibfnamefont {P.}~\bibnamefont {Sierant}}\ and\ \bibinfo {author} {\bibfnamefont {J.}~\bibnamefont {Zakrzewski}},\ }\bibfield  {title} {\bibinfo {title} {Challenges to observation of many-body localization},\ }\href@noop {} {\bibfield  {journal} {\bibinfo  {journal} {Physical Review B}\ }\textbf {\bibinfo {volume} {105}},\ \bibinfo {pages} {224203} (\bibinfo {year} {2022})}\BibitemShut {NoStop}%
\bibitem [{\citenamefont {Šuntajs}\ \emph {et~al.}(2021)\citenamefont {Šuntajs}, \citenamefont {Prosen},\ and\ \citenamefont {Vidmar}}]{Suntajs2021Spectral}%
  \BibitemOpen
  \bibfield  {author} {\bibinfo {author} {\bibfnamefont {J.}~\bibnamefont {Šuntajs}}, \bibinfo {author} {\bibfnamefont {T.}~\bibnamefont {Prosen}},\ and\ \bibinfo {author} {\bibfnamefont {L.}~\bibnamefont {Vidmar}},\ }\bibfield  {title} {\bibinfo {title} {{Spectral properties of three-dimensional Anderson model}},\ }\href {https://doi.org/https://doi.org/10.1016/j.aop.2021.168469} {\bibfield  {journal} {\bibinfo  {journal} {Annals of Physics}\ }\textbf {\bibinfo {volume} {435}},\ \bibinfo {pages} {168469} (\bibinfo {year} {2021})},\ \bibinfo {note} {special issue on Philip W. Anderson}\BibitemShut {NoStop}%
\bibitem [{\citenamefont {Sels}\ and\ \citenamefont {Polkovnikov}(2021)}]{Sels2021Dynamical}%
  \BibitemOpen
  \bibfield  {author} {\bibinfo {author} {\bibfnamefont {D.}~\bibnamefont {Sels}}\ and\ \bibinfo {author} {\bibfnamefont {A.}~\bibnamefont {Polkovnikov}},\ }\bibfield  {title} {\bibinfo {title} {{Dynamical obstruction to localization in a disordered spin chain}},\ }\href {https://doi.org/10.1103/PhysRevE.104.054105} {\bibfield  {journal} {\bibinfo  {journal} {Phys. Rev. E}\ }\textbf {\bibinfo {volume} {104}},\ \bibinfo {pages} {054105} (\bibinfo {year} {2021})}\BibitemShut {NoStop}%
\bibitem [{\citenamefont {Crowley}\ and\ \citenamefont {Chandran}(2022{\natexlab{a}})}]{Crowley2022Constructive}%
  \BibitemOpen
  \bibfield  {author} {\bibinfo {author} {\bibfnamefont {P.~J.~D.}\ \bibnamefont {Crowley}}\ and\ \bibinfo {author} {\bibfnamefont {A.}~\bibnamefont {Chandran}},\ }\bibfield  {title} {\bibinfo {title} {{A constructive theory of the numerically accessible many-body localized to thermal crossover}},\ }\href {https://doi.org/10.21468/SciPostPhys.12.6.201} {\bibfield  {journal} {\bibinfo  {journal} {SciPost Phys.}\ }\textbf {\bibinfo {volume} {12}},\ \bibinfo {pages} {201} (\bibinfo {year} {2022}{\natexlab{a}})}\BibitemShut {NoStop}%
\bibitem [{\citenamefont {Abrahams}\ \emph {et~al.}(1979)\citenamefont {Abrahams}, \citenamefont {Anderson}, \citenamefont {Licciardello},\ and\ \citenamefont {Ramakrishnan}}]{abrahams1979scaling}%
  \BibitemOpen
  \bibfield  {author} {\bibinfo {author} {\bibfnamefont {E.}~\bibnamefont {Abrahams}}, \bibinfo {author} {\bibfnamefont {P.~W.}\ \bibnamefont {Anderson}}, \bibinfo {author} {\bibfnamefont {D.~C.}\ \bibnamefont {Licciardello}},\ and\ \bibinfo {author} {\bibfnamefont {T.~V.}\ \bibnamefont {Ramakrishnan}},\ }\bibfield  {title} {\bibinfo {title} {Scaling theory of localization: Absence of quantum diffusion in two dimensions},\ }\href {https://doi.org/10.1103/PhysRevLett.42.673} {\bibfield  {journal} {\bibinfo  {journal} {Phys. Rev. Lett.}\ }\textbf {\bibinfo {volume} {42}},\ \bibinfo {pages} {673} (\bibinfo {year} {1979})}\BibitemShut {NoStop}%
\bibitem [{\citenamefont {Ma}\ \emph {et~al.}(1979)\citenamefont {Ma}, \citenamefont {Dasgupta},\ and\ \citenamefont {Hu}}]{Ma1979Random}%
  \BibitemOpen
  \bibfield  {author} {\bibinfo {author} {\bibfnamefont {S.-K.}\ \bibnamefont {Ma}}, \bibinfo {author} {\bibfnamefont {C.}~\bibnamefont {Dasgupta}},\ and\ \bibinfo {author} {\bibfnamefont {C.-K.}\ \bibnamefont {Hu}},\ }\bibfield  {title} {\bibinfo {title} {Random antiferromagnetic chain},\ }\href {https://doi.org/10.1103/PhysRevLett.43.1434} {\bibfield  {journal} {\bibinfo  {journal} {Phys. Rev. Lett.}\ }\textbf {\bibinfo {volume} {43}},\ \bibinfo {pages} {1434} (\bibinfo {year} {1979})}\BibitemShut {NoStop}%
\bibitem [{\citenamefont {Dasgupta}\ and\ \citenamefont {Ma}(1980)}]{Dasgupta1980Low}%
  \BibitemOpen
  \bibfield  {author} {\bibinfo {author} {\bibfnamefont {C.}~\bibnamefont {Dasgupta}}\ and\ \bibinfo {author} {\bibfnamefont {S.-K.}\ \bibnamefont {Ma}},\ }\bibfield  {title} {\bibinfo {title} {{Low-temperature properties of the random Heisenberg antiferromagnetic chain}},\ }\href {https://doi.org/10.1103/PhysRevB.22.1305} {\bibfield  {journal} {\bibinfo  {journal} {Phys. Rev. B}\ }\textbf {\bibinfo {volume} {22}},\ \bibinfo {pages} {1305} (\bibinfo {year} {1980})}\BibitemShut {NoStop}%
\bibitem [{\citenamefont {Fisher}(1992)}]{Fisher1992Random}%
  \BibitemOpen
  \bibfield  {author} {\bibinfo {author} {\bibfnamefont {D.~S.}\ \bibnamefont {Fisher}},\ }\bibfield  {title} {\bibinfo {title} {{Random transverse field Ising spin chains}},\ }\href {https://doi.org/10.1103/PhysRevLett.69.534} {\bibfield  {journal} {\bibinfo  {journal} {Phys. Rev. Lett.}\ }\textbf {\bibinfo {volume} {69}},\ \bibinfo {pages} {534} (\bibinfo {year} {1992})}\BibitemShut {NoStop}%
\bibitem [{\citenamefont {Fisher}(1994)}]{Fisher1994Random}%
  \BibitemOpen
  \bibfield  {author} {\bibinfo {author} {\bibfnamefont {D.~S.}\ \bibnamefont {Fisher}},\ }\bibfield  {title} {\bibinfo {title} {{Random antiferromagnetic quantum spin chains}},\ }\href {https://doi.org/10.1103/PhysRevB.50.3799} {\bibfield  {journal} {\bibinfo  {journal} {Phys. Rev. B}\ }\textbf {\bibinfo {volume} {50}},\ \bibinfo {pages} {3799} (\bibinfo {year} {1994})}\BibitemShut {NoStop}%
\bibitem [{\citenamefont {Fisher}(1995)}]{Fisher1995Critical}%
  \BibitemOpen
  \bibfield  {author} {\bibinfo {author} {\bibfnamefont {D.~S.}\ \bibnamefont {Fisher}},\ }\bibfield  {title} {\bibinfo {title} {{Critical behavior of random transverse-field Ising spin chains}},\ }\href {https://doi.org/10.1103/PhysRevB.51.6411} {\bibfield  {journal} {\bibinfo  {journal} {Phys. Rev. B}\ }\textbf {\bibinfo {volume} {51}},\ \bibinfo {pages} {6411} (\bibinfo {year} {1995})}\BibitemShut {NoStop}%
\bibitem [{\citenamefont {Vosk}\ and\ \citenamefont {Altman}(2013)}]{Vosk2013Many}%
  \BibitemOpen
  \bibfield  {author} {\bibinfo {author} {\bibfnamefont {R.}~\bibnamefont {Vosk}}\ and\ \bibinfo {author} {\bibfnamefont {E.}~\bibnamefont {Altman}},\ }\bibfield  {title} {\bibinfo {title} {Many-body localization in one dimension as a dynamical renormalization group fixed point},\ }\href {https://doi.org/10.1103/PhysRevLett.110.067204} {\bibfield  {journal} {\bibinfo  {journal} {Phys. Rev. Lett.}\ }\textbf {\bibinfo {volume} {110}},\ \bibinfo {pages} {067204} (\bibinfo {year} {2013})}\BibitemShut {NoStop}%
\bibitem [{\citenamefont {Pekker}\ \emph {et~al.}(2014)\citenamefont {Pekker}, \citenamefont {Refael}, \citenamefont {Altman}, \citenamefont {Demler},\ and\ \citenamefont {Oganesyan}}]{Pekker2014Hilbert}%
  \BibitemOpen
  \bibfield  {author} {\bibinfo {author} {\bibfnamefont {D.}~\bibnamefont {Pekker}}, \bibinfo {author} {\bibfnamefont {G.}~\bibnamefont {Refael}}, \bibinfo {author} {\bibfnamefont {E.}~\bibnamefont {Altman}}, \bibinfo {author} {\bibfnamefont {E.}~\bibnamefont {Demler}},\ and\ \bibinfo {author} {\bibfnamefont {V.}~\bibnamefont {Oganesyan}},\ }\bibfield  {title} {\bibinfo {title} {Hilbert-glass transition: New universality of temperature-tuned many-body dynamical quantum criticality},\ }\href {https://doi.org/10.1103/PhysRevX.4.011052} {\bibfield  {journal} {\bibinfo  {journal} {Phys. Rev. X}\ }\textbf {\bibinfo {volume} {4}},\ \bibinfo {pages} {011052} (\bibinfo {year} {2014})}\BibitemShut {NoStop}%
\bibitem [{\citenamefont {Potter}\ \emph {et~al.}(2015)\citenamefont {Potter}, \citenamefont {Vasseur},\ and\ \citenamefont {Parameswaran}}]{Potter2015Universal}%
  \BibitemOpen
  \bibfield  {author} {\bibinfo {author} {\bibfnamefont {A.~C.}\ \bibnamefont {Potter}}, \bibinfo {author} {\bibfnamefont {R.}~\bibnamefont {Vasseur}},\ and\ \bibinfo {author} {\bibfnamefont {S.~A.}\ \bibnamefont {Parameswaran}},\ }\bibfield  {title} {\bibinfo {title} {Universal properties of many-body delocalization transitions},\ }\href {https://doi.org/10.1103/PhysRevX.5.031033} {\bibfield  {journal} {\bibinfo  {journal} {Phys. Rev. X}\ }\textbf {\bibinfo {volume} {5}},\ \bibinfo {pages} {031033} (\bibinfo {year} {2015})}\BibitemShut {NoStop}%
\bibitem [{\citenamefont {Zhang}\ \emph {et~al.}(2016)\citenamefont {Zhang}, \citenamefont {Zhao}, \citenamefont {Devakul},\ and\ \citenamefont {Huse}}]{Zhang2016Many}%
  \BibitemOpen
  \bibfield  {author} {\bibinfo {author} {\bibfnamefont {L.}~\bibnamefont {Zhang}}, \bibinfo {author} {\bibfnamefont {B.}~\bibnamefont {Zhao}}, \bibinfo {author} {\bibfnamefont {T.}~\bibnamefont {Devakul}},\ and\ \bibinfo {author} {\bibfnamefont {D.~A.}\ \bibnamefont {Huse}},\ }\bibfield  {title} {\bibinfo {title} {{Many-body localization phase transition: A simplified strong-randomness approximate renormalization group}},\ }\href {https://doi.org/10.1103/PhysRevB.93.224201} {\bibfield  {journal} {\bibinfo  {journal} {Phys. Rev. B}\ }\textbf {\bibinfo {volume} {93}},\ \bibinfo {pages} {224201} (\bibinfo {year} {2016})}\BibitemShut {NoStop}%
\bibitem [{\citenamefont {Thiery}\ \emph {et~al.}(2017)\citenamefont {Thiery}, \citenamefont {M\"uller},\ and\ \citenamefont {De~Roeck}}]{Thiery2017Microscopically}%
  \BibitemOpen
  \bibfield  {author} {\bibinfo {author} {\bibfnamefont {T.}~\bibnamefont {Thiery}}, \bibinfo {author} {\bibfnamefont {M.}~\bibnamefont {M\"uller}},\ and\ \bibinfo {author} {\bibfnamefont {W.}~\bibnamefont {De~Roeck}},\ }\bibfield  {title} {\bibinfo {title} {{A microscopically motivated renormalization scheme for the MBL/ETH transition}},\ }\href {https://arxiv.org/abs/1711.09880} {\bibfield  {journal} {\bibinfo  {journal} {arXiv:1711.09880}\ } (\bibinfo {year} {2017})}\BibitemShut {NoStop}%
\bibitem [{\citenamefont {Cardy}(1988)}]{Cardy1988}%
  \BibitemOpen
  \bibinfo {editor} {\bibfnamefont {J.}~\bibnamefont {Cardy}},\ ed.,\ \href@noop {} {\emph {\bibinfo {title} {Finite-Size Scaling}}},\ \bibinfo {number} {2}\ (\bibinfo  {publisher} {North Holland},\ \bibinfo {year} {1988})\BibitemShut {NoStop}%
\bibitem [{\citenamefont {Pelissetto}\ and\ \citenamefont {Vicari}(2002)}]{PELISSETTO_RG}%
  \BibitemOpen
  \bibfield  {author} {\bibinfo {author} {\bibfnamefont {A.}~\bibnamefont {Pelissetto}}\ and\ \bibinfo {author} {\bibfnamefont {E.}~\bibnamefont {Vicari}},\ }\bibfield  {title} {\bibinfo {title} {{Critical phenomena and renormalization-group theory}},\ }\href {https://www.sciencedirect.com/science/article/pii/S0370157302002193} {\bibfield  {journal} {\bibinfo  {journal} {Physics Reports}\ }\textbf {\bibinfo {volume} {368}},\ \bibinfo {pages} {549} (\bibinfo {year} {2002})}\BibitemShut {NoStop}%
\bibitem [{Note1()}]{Note1}%
  \BibitemOpen
  \bibinfo {note} {It is possible to work not strictly in the limit $\omega \to \infty $, and analyze the flow close to the critical point to extract the correct relevant and irrelevant anomalous dimensions to $O(1/\omega )$.}\BibitemShut {Stop}%
\bibitem [{\citenamefont {Sierant}\ \emph {et~al.}(2023{\natexlab{a}})\citenamefont {Sierant}, \citenamefont {Lewenstein},\ and\ \citenamefont {Scardicchio}}]{sierant2023universality}%
  \BibitemOpen
  \bibfield  {author} {\bibinfo {author} {\bibfnamefont {P.}~\bibnamefont {Sierant}}, \bibinfo {author} {\bibfnamefont {M.}~\bibnamefont {Lewenstein}},\ and\ \bibinfo {author} {\bibfnamefont {A.}~\bibnamefont {Scardicchio}},\ }\bibfield  {title} {\bibinfo {title} {Universality in anderson localization on random graphs with varying connectivity},\ }\href@noop {} {\bibfield  {journal} {\bibinfo  {journal} {SciPost Physics}\ }\textbf {\bibinfo {volume} {15}},\ \bibinfo {pages} {045} (\bibinfo {year} {2023}{\natexlab{a}})}\BibitemShut {NoStop}%
\bibitem [{\citenamefont {Sierant}\ \emph {et~al.}(2023{\natexlab{b}})\citenamefont {Sierant}, \citenamefont {Lewenstein},\ and\ \citenamefont {Scardicchio}}]{sierant23a}%
  \BibitemOpen
  \bibfield  {author} {\bibinfo {author} {\bibfnamefont {P.}~\bibnamefont {Sierant}}, \bibinfo {author} {\bibfnamefont {M.}~\bibnamefont {Lewenstein}},\ and\ \bibinfo {author} {\bibfnamefont {A.}~\bibnamefont {Scardicchio}},\ }\bibfield  {title} {\bibinfo {title} {{Universality in Anderson localization on random graphs with varying connectivity}},\ }\href {https://doi.org/10.21468/SciPostPhys.15.2.045} {\bibfield  {journal} {\bibinfo  {journal} {SciPost Phys.}\ }\textbf {\bibinfo {volume} {15}},\ \bibinfo {pages} {045} (\bibinfo {year} {2023}{\natexlab{b}})}\BibitemShut {NoStop}%
\bibitem [{\citenamefont {Harris}(1974)}]{Harris1974Effect}%
  \BibitemOpen
  \bibfield  {author} {\bibinfo {author} {\bibfnamefont {A.~B.}\ \bibnamefont {Harris}},\ }\bibfield  {title} {\bibinfo {title} {Effect of random defects on the critical behaviour of ising models},\ }\href {https://doi.org/10.1088/0022-3719/7/9/009} {\bibfield  {journal} {\bibinfo  {journal} {J. Phys. C}\ }\textbf {\bibinfo {volume} {7}},\ \bibinfo {pages} {1671} (\bibinfo {year} {1974})}\BibitemShut {NoStop}%
\bibitem [{\citenamefont {Cardy}(1996)}]{Cardy_1996}%
  \BibitemOpen
  \bibfield  {author} {\bibinfo {author} {\bibfnamefont {J.}~\bibnamefont {Cardy}},\ }\href@noop {} {\emph {\bibinfo {title} {{Scaling and Renormalization in Statistical Physics}}}},\ Cambridge Lecture Notes in Physics\ (\bibinfo  {publisher} {Cambridge University Press},\ \bibinfo {year} {1996})\BibitemShut {NoStop}%
\bibitem [{Note2()}]{Note2}%
  \BibitemOpen
  \bibinfo {note} {The logarithmic derivatives in Eq.~~\protect \eqref {eq:1p_scaling} are computed by using $N=2^L$ in place of $N = \protect \binom {L}{L/2}$, as they differ by a proportionality constant.}\BibitemShut {Stop}%
\bibitem [{\citenamefont {Abou-Chacra}\ \emph {et~al.}(1973)\citenamefont {Abou-Chacra}, \citenamefont {Thouless},\ and\ \citenamefont {Anderson}}]{abou1973selfconsistent}%
  \BibitemOpen
  \bibfield  {author} {\bibinfo {author} {\bibfnamefont {R.}~\bibnamefont {Abou-Chacra}}, \bibinfo {author} {\bibfnamefont {D.}~\bibnamefont {Thouless}},\ and\ \bibinfo {author} {\bibfnamefont {P.}~\bibnamefont {Anderson}},\ }\bibfield  {title} {\bibinfo {title} {A selfconsistent theory of localization},\ }\href@noop {} {\bibfield  {journal} {\bibinfo  {journal} {J. Phys. C: Solid State Physics}\ }\textbf {\bibinfo {volume} {6}},\ \bibinfo {pages} {1734} (\bibinfo {year} {1973})}\BibitemShut {NoStop}%
\bibitem [{\citenamefont {Parisi}\ \emph {et~al.}(2019)\citenamefont {Parisi}, \citenamefont {Pascazio}, \citenamefont {Pietracaprina}, \citenamefont {Ros},\ and\ \citenamefont {Scardicchio}}]{parisi2019anderson}%
  \BibitemOpen
  \bibfield  {author} {\bibinfo {author} {\bibfnamefont {G.}~\bibnamefont {Parisi}}, \bibinfo {author} {\bibfnamefont {S.}~\bibnamefont {Pascazio}}, \bibinfo {author} {\bibfnamefont {F.}~\bibnamefont {Pietracaprina}}, \bibinfo {author} {\bibfnamefont {V.}~\bibnamefont {Ros}},\ and\ \bibinfo {author} {\bibfnamefont {A.}~\bibnamefont {Scardicchio}},\ }\bibfield  {title} {\bibinfo {title} {Anderson transition on the bethe lattice: an approach with real energies},\ }\href {https://doi.org/10.1088/1751-8121/ab56e8} {\bibfield  {journal} {\bibinfo  {journal} {J. Phys. A: Math. Theor.}\ }\textbf {\bibinfo {volume} {53}},\ \bibinfo {pages} {014003} (\bibinfo {year} {2019})}\BibitemShut {NoStop}%
\bibitem [{\citenamefont {De~Luca}\ \emph {et~al.}(2014)\citenamefont {De~Luca}, \citenamefont {Altshuler}, \citenamefont {Kravtsov},\ and\ \citenamefont {Scardicchio}}]{de2014anderson}%
  \BibitemOpen
  \bibfield  {author} {\bibinfo {author} {\bibfnamefont {A.}~\bibnamefont {De~Luca}}, \bibinfo {author} {\bibfnamefont {B.}~\bibnamefont {Altshuler}}, \bibinfo {author} {\bibfnamefont {V.}~\bibnamefont {Kravtsov}},\ and\ \bibinfo {author} {\bibfnamefont {A.}~\bibnamefont {Scardicchio}},\ }\bibfield  {title} {\bibinfo {title} {Anderson localization on the bethe lattice: Nonergodicity of extended states},\ }\href@noop {} {\bibfield  {journal} {\bibinfo  {journal} {Physical review letters}\ }\textbf {\bibinfo {volume} {113}},\ \bibinfo {pages} {046806} (\bibinfo {year} {2014})}\BibitemShut {NoStop}%
\bibitem [{\citenamefont {Bogomolny}\ \emph {et~al.}(2011)\citenamefont {Bogomolny}, \citenamefont {Giraud},\ and\ \citenamefont {Schmit}}]{Bogomolny2011Integrable}%
  \BibitemOpen
  \bibfield  {author} {\bibinfo {author} {\bibfnamefont {E.}~\bibnamefont {Bogomolny}}, \bibinfo {author} {\bibfnamefont {O.}~\bibnamefont {Giraud}},\ and\ \bibinfo {author} {\bibfnamefont {C.}~\bibnamefont {Schmit}},\ }\bibfield  {title} {\bibinfo {title} {Integrable random matrix ensembles},\ }\href {https://doi.org/10.1088/0951-7715/24/11/010} {\bibfield  {journal} {\bibinfo  {journal} {Nonlinearity}\ }\textbf {\bibinfo {volume} {24}},\ \bibinfo {pages} {3179–3213} (\bibinfo {year} {2011})}\BibitemShut {NoStop}%
\bibitem [{\citenamefont {Laumann}\ \emph {et~al.}(2014)\citenamefont {Laumann}, \citenamefont {Pal},\ and\ \citenamefont {Scardicchio}}]{Laumann2014MBMobility}%
  \BibitemOpen
  \bibfield  {author} {\bibinfo {author} {\bibfnamefont {C.~R.}\ \bibnamefont {Laumann}}, \bibinfo {author} {\bibfnamefont {A.}~\bibnamefont {Pal}},\ and\ \bibinfo {author} {\bibfnamefont {A.}~\bibnamefont {Scardicchio}},\ }\bibfield  {title} {\bibinfo {title} {Many-body mobility edge in a mean-field quantum spin glass},\ }\href {https://doi.org/10.1103/PhysRevLett.113.200405} {\bibfield  {journal} {\bibinfo  {journal} {Phys. Rev. Lett.}\ }\textbf {\bibinfo {volume} {113}},\ \bibinfo {pages} {200405} (\bibinfo {year} {2014})}\BibitemShut {NoStop}%
\bibitem [{\citenamefont {Ponte}\ \emph {et~al.}(2017)\citenamefont {Ponte}, \citenamefont {Laumann}, \citenamefont {Huse},\ and\ \citenamefont {Chandran}}]{ponte2017thermal}%
  \BibitemOpen
  \bibfield  {author} {\bibinfo {author} {\bibfnamefont {P.}~\bibnamefont {Ponte}}, \bibinfo {author} {\bibfnamefont {C.}~\bibnamefont {Laumann}}, \bibinfo {author} {\bibfnamefont {D.~A.}\ \bibnamefont {Huse}},\ and\ \bibinfo {author} {\bibfnamefont {A.}~\bibnamefont {Chandran}},\ }\bibfield  {title} {\bibinfo {title} {Thermal inclusions: how one spin can destroy a many-body localized phase},\ }\href@noop {} {\bibfield  {journal} {\bibinfo  {journal} {Philosophical Transactions of the Royal Society A: Mathematical, Physical and Engineering Sciences}\ }\textbf {\bibinfo {volume} {375}},\ \bibinfo {pages} {20160428} (\bibinfo {year} {2017})}\BibitemShut {NoStop}%
\bibitem [{\citenamefont {\ifmmode~\check{S}\else \v{S}\fi{}untajs}\ and\ \citenamefont {Vidmar}(2022)}]{suntajs_vidmar_22}%
  \BibitemOpen
  \bibfield  {author} {\bibinfo {author} {\bibfnamefont {J.}~\bibnamefont {\ifmmode~\check{S}\else \v{S}\fi{}untajs}}\ and\ \bibinfo {author} {\bibfnamefont {L.}~\bibnamefont {Vidmar}},\ }\bibfield  {title} {\bibinfo {title} {Ergodicity breaking transition in zero dimensions},\ }\href {https://doi.org/10.1103/PhysRevLett.129.060602} {\bibfield  {journal} {\bibinfo  {journal} {Phys. Rev. Lett.}\ }\textbf {\bibinfo {volume} {129}},\ \bibinfo {pages} {060602} (\bibinfo {year} {2022})}\BibitemShut {NoStop}%
\bibitem [{\citenamefont {Crowley}\ and\ \citenamefont {Chandran}(2022{\natexlab{b}})}]{crowley2022partial}%
  \BibitemOpen
  \bibfield  {author} {\bibinfo {author} {\bibfnamefont {P.}~\bibnamefont {Crowley}}\ and\ \bibinfo {author} {\bibfnamefont {A.}~\bibnamefont {Chandran}},\ }\bibfield  {title} {\bibinfo {title} {Partial thermalisation of a two-state system coupled to a finite quantum bath},\ }\href@noop {} {\bibfield  {journal} {\bibinfo  {journal} {SciPost Physics}\ }\textbf {\bibinfo {volume} {12}},\ \bibinfo {pages} {103} (\bibinfo {year} {2022}{\natexlab{b}})}\BibitemShut {NoStop}%
\bibitem [{\citenamefont {Balay}\ \emph {et~al.}(2024)\citenamefont {Balay}, \citenamefont {Abhyankar}, \citenamefont {Adams} \emph {et~al.}}]{PETSc}%
  \BibitemOpen
  \bibfield  {author} {\bibinfo {author} {\bibfnamefont {S.}~\bibnamefont {Balay}}, \bibinfo {author} {\bibfnamefont {S.}~\bibnamefont {Abhyankar}}, \bibinfo {author} {\bibfnamefont {M.~F.}\ \bibnamefont {Adams}}, \emph {et~al.},\ }\href {https://petsc.org/} {\bibinfo {title} {{PETS}c {W}eb page}},\ \bibinfo {howpublished} {\url{https://petsc.org/}} (\bibinfo {year} {2024})\BibitemShut {NoStop}%
\bibitem [{\citenamefont {Hernandez}\ \emph {et~al.}(2005)\citenamefont {Hernandez}, \citenamefont {Roman},\ and\ \citenamefont {Vidal}}]{SLEPc}%
  \BibitemOpen
  \bibfield  {author} {\bibinfo {author} {\bibfnamefont {V.}~\bibnamefont {Hernandez}}, \bibinfo {author} {\bibfnamefont {J.~E.}\ \bibnamefont {Roman}},\ and\ \bibinfo {author} {\bibfnamefont {V.}~\bibnamefont {Vidal}},\ }\bibfield  {title} {\bibinfo {title} {{SLEPc: A scalable and flexible toolkit for the solution of eigenvalue problems}},\ }\href {https://doi.org/10.1145/1089014.1089019} {\bibfield  {journal} {\bibinfo  {journal} {ACM Trans. Math. Softw.}\ }\textbf {\bibinfo {volume} {31}},\ \bibinfo {pages} {351–362} (\bibinfo {year} {2005})}\BibitemShut {NoStop}%
\bibitem [{\citenamefont {ReCaS-Bari}()}]{ReCaS}%
  \BibitemOpen
  \bibfield  {author} {\bibinfo {author} {\bibnamefont {ReCaS-Bari}},\ }\href@noop {} {\bibinfo {title} {https://www.recas-bari.it/}}\BibitemShut {NoStop}%
\end{thebibliography}%

\end{document}